\documentclass[12pt]{elsarticle}
\usepackage{amssymb}
\usepackage{amsmath}
\usepackage{amsthm}
\usepackage{subfigure}
\usepackage{graphicx}

\biboptions{numbers,sort&compress}

\usepackage[dvipdfm,colorlinks,linkcolor=red,anchorcolor=blue,citecolor=blue]{hyperref}
\usepackage{booktabs}
\makeatletter

\newcommand{\Rmnum}[1]{\expandafter\@slowromancap\romannumeral #1@}
\makeatother
\usepackage{caption}

\usepackage{anysize}
\usepackage{setspace}
\marginsize{2.0cm}{1.5cm}{0.8cm}{0.8cm}
\journal{Journal of XX}

\begin{document}

\begin{frontmatter}

\title{Modulation instability and controllable rogue waves with multiple compression points for periodically
modulated coupled Hirota equations }
\author{ Xin Wang }
\author{ Yong Chen$^{*}$}\ead{ychen@sei.ecnu.edu.cn}

\cortext[cor1]{Corresponding author.}

\address{Shanghai Key Laboratory of Trustworthy Computing, East China Normal University, Shanghai, 200062,
People's Republic of China}

\begin{abstract}
Based on modulation instability analysis and generalized Darboux transformation,
we derive a hierarchy of rogue wave solutions for a variable-coefficients coupled Hirota equations.
The explicit first-order rogue wave solution is presented, and
the dark-bright and composite rogue waves with multiple compression points are shown
by choosing sufficiently large periodic modulation amplitudes in the coefficients of the coupled equations.
Also, the dark-bright and composite Peregrine combs are generated from the multiple compression points.
For the second-order case,
the dark-bright three sisters, rogue wave quartets and sextets structures with one or more rogue waves involving
multiple compression points are put forward, respectively.
Furthermore, some wave characteristics such as the
difference between light intensity and continuous wave background,
and pulse energy evolution of the dark rogue wave solution features multiple compression points are discussed.
\end{abstract}


\end{frontmatter}

\section{Introduction }  

Originally termed to describe monstrous sea wave events
in oceanography \cite{01,02}, rogue waves have been at the center of considerable research activity
to a great extent due to their emerging relevance in a great variety of realms. These includes,
the nonlinear optics \cite{03}, Bose-Einstein condensates (BEC) \cite{04}, atmosphere \cite{05},
surface plasma \cite{06}, and even econophysics \cite{07}.
They suddenly emerge with an amplitude significantly larger than that of the
surrounding wave crests and vanish without slighting trace, and can be generally
expressed by rational polynomials in mathematics \cite{08}.

The foremost description of a single rogue wave is the Peregrine soliton \cite{09},
a rational solution of the nonlinear Schr\"{o}dinger (NLS) equation which features
a localized peak whose amplitude is three times larger than that of the average height.
Subsequently, the more complicated rogue waves which can be represented
by higher-order rational functions have been systematically investigated for the NLS equation \cite{10,11,12}.
Additionally, recent experiments in a water tank indicate that
the actual dynamics of these extreme waves can be commendably described by the analytic solutions \cite{13,14}.

Considering the various physical contexts, one should go beyond the standard NLS representation
to make further efforts to reveal the complex dynamics of rogue waves.
Therefore, some important generalized higher-order systems (e.g. the Hirota equation \cite{15},
quartic NLS equations \cite{16,17}  and
Kundu-Eckhaus equation \cite{18}) and coupled systems (e.g. the Manakov system \cite{19,20,21},
coupled Hirota equations \cite{22,23,24} and three-wave resonant interaction equations \cite{25})
have been paid widespread concern.
In fact, it is confirmed that the higher-order dispersion and/or nonlinear terms play
a pivotal pole in affecting the dynamics of rogue waves, including their \lq ridge' \cite{15} and
temporal-spacial distributions \cite{26}, etc.
Especially, the very recent reports on \lq breather-to-soliton' and \lq transition' properties
show that the interesting w-shaped soliton can exist in the Hirota equation instead of the standard
NLS model without higher-order effects \cite{27,28}. In addition, it is recognized that rogue waves
in coupled systems can display diversity and complexity, which contains, the rogue wave-breather/soliton
interactional structure \cite{19,24},  dark structure \cite{20},
composite structure \cite{21}, four-petaled structure \cite{29}, and so forth.

Management of rogue waves in inhomogeneous BEC has also received extensive attention within the past decade \cite{30,31,32,33,34}.
A surge of work have been reported on involving controlling rogue waves in nonautonomous single or multi-components
NLS equations with space and/or time modulated potentials like periodic potentials, harmonic potential, to name a few \cite{30}.
In particular, very recently, Tiofack et al.  study the comb generation of the
periodically modulated NLS equation  by using the multiple compression points approach,
which paves an important way for the experimental control and manipulation of nonautonomous rogue waves modeled by the
NLS equation \cite{35}.

The propagation of vector optical rogue waves with higher-order effects in inhomogeneous optical
fibers is described by the variable-coefficients coupled Hirota (VCCH) equations \cite{36,37},
which can be written as the following dimensionless form:
\begin{align}
&{\rm i}u_{1t}+\alpha(t)u_{1xx}+2\delta(t)(|u_{1}|^{2}+|u_{2}|^{2})u_{1}+{\rm i}\beta(t)[u_{1xxx}+\gamma(t)(6|u_{1}|^{2}
+3|u_{2}|^{2})u_{1x}+3\gamma(t)u_{1}u_{2}^{*}u_{2x}]\nonumber\\&~~~~~+{\rm i}\Gamma(t)u_{1}=0, \label{01}\\
&{\rm i}u_{2t}+\alpha(t)u_{2xx}+2\delta(t)(|u_{1}|^{2}+|u_{2}|^{2})u_{2}+{\rm i}\beta(t)[u_{2xxx}+\gamma(t)(6|u_{2}|^{2}
+3|u_{1}|^{2})u_{2x}+3\gamma(t)u_{2}u_{1}^{*}u_{1x}]\nonumber\\&~~~~~+{\rm i}\Gamma(t)u_{2}=0,\label{02}
\end{align}
where
$$
\gamma(t)=\frac{\delta(t)}{\alpha(t)},\ \Gamma(t)=\frac{1}{2}\bigg[\frac{\delta(t)_{t}}{\delta(t)}-\frac{\alpha(t)_{t}}{\alpha(t)}\bigg]
$$
is satisfied to ensure the Painlev\'{e} integrability of Eqs. (\ref{01})-(\ref{02}). The potentials
$u_{1}(x,t)$, $u_{2}(x,t)$ are the complex wave envelops, $t$ and $x$ represent the
propagation distance and transverse coordinate, and asterisk means complex conjugate.
$\alpha(t)$, $\beta(t)$, $\delta(t)$, $\gamma(t)$ and $\Gamma(t)$
are restricted to periodic functions and
stand for group velocity dispersion (GVD), third-order dispersion (TOD), nonlinear terms
referred to self-phase modulation (SPM) and cross-phase modulation (XPM), nonlinear terms related to
self-steepening (SS) and stimulated Raman scattering (SRS), and gain or absorption modulus, respectively.

In the present work, we construct the Lax pair for Eqs. (\ref{01})-(\ref{02}) by making use of the AKNS technique.
Next based on modulation instability (MI) analysis \cite{38} and
generalized Darboux transformation (gDT) \cite{39,40,41,42,43},
we derive a hierarchy of general rogue wave solutions for Eqs. (\ref{01})-(\ref{02}).
Explicit first-order rogue wave solution  is presented,
the dark-bright and composite rogue waves with multiple compression points are shown
under the sinusoidally varying coefficients modulated conditions.
Then the dark-bright and composite Peregrine combs are generated from the multiple compression points,
as long as the amplitude of the periodic modulation is sufficiently largely chosen.
For the second-order circumstance, the dark-bright three sisters, rogue wave quartets and sextets
structures containing one or more rogue waves with
multiple compression points are displayed, respectively.
The rest of this paper discuss some important wave characteristics such as the
difference between light intensity and continuous-wave (CW) background,
and pulse energy evolution to the dark rogue wave solution features multiple compression points.

The paper is structured as follows. In section 2, the linear stability
of a CW solution regarding to MI is analyzed. In section 3,
the Lax pair, gDT and a hierarchy of general rogue wave solutions are derived.
In section 4, rogue waves with multiple compression points are studied. In section 5,
further properties of the dark rogue wave solution are discussed. In section 6, we
summarize our results and provide some discussions.

\section{Modulation instability}
Our starting point is a CW solution of Eqs. (\ref{01})-(\ref{02}) with the generalized form
\begin{equation}\label{03}
u_{1}[0]=\sqrt{\frac{\alpha(t)}{\delta(t)}}ce^{{\rm i}\theta_{1}},\
u_{2}[0]=\sqrt{\frac{\alpha(t)}{\delta(t)}}ce^{{\rm i}\theta_{2}},
\end{equation}
where $\theta_{i}=a_{i}x+b_{i}(t)(i=1,2)$,
$c$ and $a_{i}$ are real constants and delegate the amplitude and frequency of the CW background, and
$b_{i}(t)$ will be determined in the following calculations.

It is known that frequency difference between two modes can produce
significant physical effects, and it has been recently proved that the frequency relationship
$a_{1}-a_{2}=2a$ could bring about abundant types of rogue waves in coupled systems (see \cite{21,22,28} and the references therein).
Accordingly, we now uniformly set the CW fields with different frequencies (i.e. $a_{1}=-a_{2}=a$).
At this point, the function $b_{i}(t)$ can be determined by
\begin{align}
&b_{1}(t)=(4c^{2}-a^{2})A(t)+(a^3-6ac^{2})B(t),\label{04}\\
&b_{2}(t)=(4c^{2}-a^{2})A(t)-(a^3-6ac^{2})B(t),\label{05}
\end{align}
where
\begin{equation}\label{AB}
A(t)=\int^{t} \alpha(t')\mathrm{d}t'+A_{0},\ B(t)=\int^{t} \beta(t')\mathrm{d}t'+B_{0},
\end{equation}
with $A_{0}$, $B_{0}$ being two integration constants.

Next we impose a small perturbation on the CW solution by taking
\begin{equation}\label{06}
u_{1}=u_{1}[0](1+q_{1}),\ u_{2}=u_{2}[0](1+q_{2}).
\end{equation}
Substituting Eq. (\ref{06}) into Eqs. (\ref{01})-(\ref{02}) one can get the linearized VCCH equations as
\begin{align}
&{\rm i}q_{1t}+(2\alpha c^2-6\beta ac^2)(q_{1}+q_{1}^{*})+2\alpha c^2 (q_{2}+q_{2}^{*})+{\rm i}(9\beta c^2-3\beta a^2+2\alpha a)q_{1x}
+3{\rm i}\beta c^2 q_{2x}\nonumber\\&~~~~+(\alpha-3\beta a)q_{1xx}+{\rm i}\beta q_{1xxx}=0,\label{07}\\
&{\rm i}q_{2t}+(2\alpha c^2+6\beta ac^2)(q_{2}+q_{2}^{*})+2\alpha c^2 (q_{1}+q_{1}^{*})+{\rm i}(9\beta c^2-3\beta a^2-2\alpha a)q_{2x}
+3{\rm i}\beta c^2 q_{1x}\nonumber\\&~~~~+(\alpha+3\beta a)q_{2xx}+{\rm i}\beta q_{2xxx}=0.\label{08}
\end{align}
The stability of the solution of the above linearized equations to wavelength $\kappa$
can be studied by collecting the Fourier modes in the following way:
\begin{equation}\label{09}
q_{1}=f_{+}\exp[{\rm i}\kappa(x-\Omega(t))]+f_{-}^{*}\exp[{\rm i}\kappa(x-\Omega(t)^{*})],\
q_{2}=g_{+}\exp[{\rm i}\kappa(x-\Omega(t))]+g_{-}^{*}\exp[{\rm i}\kappa(x-\Omega(t)^{*})],
\end{equation}
where  $\Omega(t)$ is a non-real function corresponding to instability.
Putting Eq. (\ref{09}) into Eqs. (\ref{07})-(\ref{08}) and to ensure the equations about
$\{f_{+},f_{-},g_{+},g_{-}\}$  be solvable we have
\begin{align}
&\Omega(t)=(9c^2-3a^2-\kappa^2)B(t)+3\sqrt{c^4-4a^2c^2+\kappa^2a^2}B(t)
\nonumber\\&~~~~~~+\sqrt{4a^2-4c^2+\kappa^2-4\sqrt{c^4-4a^2c^2+\kappa^2 a^2  }}A(t).
\end{align}
Notice that the higher-order terms (i.e. TOD, SS and  SRS)  can affect the
characteristics of MI such that the condition
\begin{equation}\label{11}
c^4-4a^2c^2+\kappa^2a^2<0
\end{equation}
is fulfilled. Under this assumption, we prove that the MI exist and the corresponding growth rate can be obtained as
\begin{equation}
\begin{array}{l}
{\rm Im}\{\Omega(t)\}=\displaystyle3\sqrt{-c^4+4a^2c^2-\kappa^2a^2}B(t)
\\~~~~~~~~~~~~+\displaystyle\dfrac{1}{2}\sqrt{-2\kappa^2-8 a^2+8 c^2
+2\sqrt{(4a^2-\kappa^2)(4a^2+8c^2-\kappa^2)}}A(t).
\end{array}
\end{equation}
The above MI actually provides the evidence of occurrence of rogue waves in Eqs. (\ref{01})-(\ref{02}).
In the next section, we will derive a hierarchy of rogue wave solutions based on MI and gDT.

\section{Darboux transformation and rogue wave solutions }

Firstly, we present the Lax pair of Eqs. (\ref{01})-(\ref{02}) which can be obtained through the AKNS technique:
\begin{align}
&\Psi_{x}=U\Psi,\ U=\zeta U_{0}+U_{1},\label{13}\\
&\Psi_{t}=V\Psi,\ V=\zeta^3 V_{0}+\zeta^2 V_{1}+\zeta V_{2}+V_{3},\label{14}
\end{align}
where
$$
\begin{array}{l}
U_{0}=\begin{pmatrix}
-2{\rm i} & 0 & 0\\
0 & {\rm i} & 0\\
0 &0 &{\rm i}
\end{pmatrix},\ U_{1}=
{\rm i}\sqrt{\dfrac{\delta(t)}{\alpha(t)}}\begin{pmatrix}
0 & u_{1} & u_{2}\\
u_{1}^{*} & 0 & 0\\
u_{2}^{*} &0 &0
\end{pmatrix}, V_{0}=9{\rm i}\beta(t)U_{0},\ V_{1}=3\alpha(t)U_{0}+9\beta(t)U_{1},\\
V_{2}=3{\rm i}\sqrt{\dfrac{\delta(t)}{\alpha(t)}}
\begin{pmatrix}
\sqrt{\dfrac{\delta(t)}{\alpha(t)}}\beta(t)(|u_{1}|^2+|u_{2}|^2) & {\rm i}\beta(t)u_{1x}+\alpha(t)u_{1}
& {\rm i}\beta(t)u_{2x}+\alpha(t)u_{2}\\
-{\rm i}\beta(t)u_{1x}^{*}+\alpha(t)u_{1}^{*} &
-\sqrt{\dfrac{\delta(t)}{\alpha(t)}}\beta(t)|u_{1}|^2 & -\sqrt{\dfrac{\delta(t)}{\alpha(t)}}\beta(t)u_{1}^{*}u_{2}\\
-{\rm i}\beta(t)u_{2x}^{*}+\alpha(t)u_{2}^{*}
&-\sqrt{\dfrac{\delta(t)}{\alpha(t)}}\beta(t)u_{1}u_{2}^{*} &-\sqrt{\dfrac{\delta(t)}{\alpha(t)}}\beta(t)|u_{2}|^2
\end{pmatrix},\\
V_{3}=\begin{pmatrix}
\beta(t)(e_{1}+e_{2})+{\rm i}\delta (|u_{1}|^2+|u_{2}|^2) & \beta(t)e_{3}-\sqrt{\alpha(t)\delta(t)}u_{1x} & \beta(t)e_{4}-\sqrt{\alpha(t)\delta(t)}u_{2x}\\
-\beta(t)e_{3}^{*}+\sqrt{\alpha(t)\delta(t)}u_{1x}^{*} & -\beta(t)e_{1}-{\rm i}\delta |u_{1}|^2 & -\beta(t)e_{5}-{\rm i}\delta u_{1}^{*}u_{2}\\
-\beta(t)e_{4}^{*}+\sqrt{\alpha(t)\delta(t)}u_{2x}^{*} & \beta(t)e_{5}^{*}-{\rm i}\delta u_{1}u_{2}^{*} & -\beta(t)e_{2}-{\rm i}\delta |u_{2}|^2
\end{pmatrix},
\end{array}
$$
with
$$
\begin{array}{l}
e_{1}=-\dfrac{\delta(t)}{\alpha(t)}\bigg(u_{1}^{*}u_{1x}-u_{1}u_{1x}^{*}\bigg),\ e_{2}=-\dfrac{\delta(t)}{\alpha(t)}\bigg(u_{2}^{*}u_{2x}-u_{2}u_{2x}^{*}\bigg),\\
e_{3}=-{\rm i}\sqrt{\dfrac{\delta(t)}{\alpha(t)}}\bigg[u_{1xx}+2\dfrac{\delta(t)}{\alpha(t)}(|u_{1}|^2+|u_{2}|^2)u_{1}\bigg],\\
e_{4}=-{\rm i}\sqrt{\dfrac{\delta(t)}{\alpha(t)}}\bigg[u_{2xx}+2\dfrac{\delta(t)}{\alpha(t)}(|u_{1}|^2+|u_{2}|^2)u_{2}\bigg],\\
e_{5}=-\dfrac{\delta(t)}{\alpha(t)}\bigg(u_{1}^{*}u_{2x}-u_{2}u_{1x}^{*}\bigg).
\end{array}
$$
The compatibility of the above Lax pair straightforwardly gives rise to Eqs. (\ref{01})-(\ref{02}).
 Next the DT can be constructed in the following way.

{\bf Proposition 1} The following DT
\begin{equation}
\Psi[1]=T\Psi, T=I-\dfrac{P_{1}}{\zeta-\zeta_{1}^{*}},\ P_{1}=\Psi_{1}\bigg(\dfrac{\Psi_{1}^{\dagger}\Psi_{1}}{\zeta_{1}-\zeta_{1}^{*}} \bigg)^{-1}\Psi_{1}^{\dagger},
\end{equation}
where
$\Psi_{1}=(\psi_{1},\phi_{1},\chi_{1})^{T}$ is a
basic solution of the linear spectral problem (\ref{13})-(\ref{14}) at $\zeta=\zeta_{1}$ and the initial solution of
$u_{i}=u_{i}[0]$ ($i=1,2$),
and dagger represents Hermite conjugation, converts the linear spectral problem (\ref{13})-(\ref{14})
into a new system, that is
\begin{align}
&\Psi[1]_{x}=U[1](\zeta,u_{i}[1])\Psi[1],\\
&\Psi[1]_{t}=V[1](\zeta,u_{i}[1])\Psi[1],
\end{align}
where
\begin{align}
&u_{1}[1]=u_{1}[0]-3\sqrt{\dfrac{\alpha(t)}{\delta(t)}}\dfrac{(\zeta_{1}-\zeta_{1}^{*})}{\Psi_{1}^{\dagger}\Psi_{1}}
\psi_{1}\phi_{1}^{*},\\
&u_{2}[1]=u_{2}[0]-3\sqrt{\dfrac{\alpha(t)}{\delta(t)}}\dfrac{(\zeta_{1}-\zeta_{1}^{*})}{\Psi_{1}^{\dagger}\Psi_{1}}
\psi_{1}\chi_{1}^{*}.
\end{align}

After that, on the basis of MI, one can calculate the basic solution of the Lax pair equations (\ref{13})-(\ref{14})
with the initial potentials being selected as the CW solution (\ref{03}).

By directly substituting Eq. (\ref{03}) into Eqs. (\ref{13})-(\ref{14}), one can get the general form of the basic solution,
which reads
\begin{equation}\label{20}
\Psi=D\begin{pmatrix}
1 & 1 & 1 \\
\dfrac{c}{\xi_{1}-\zeta-a} & \dfrac{c}{\xi_{2}-\zeta-a} & \dfrac{c}{\xi_{3}-\zeta-a}\\
\dfrac{c}{\xi_{1}-\zeta+a} & \dfrac{c}{\xi_{2}-\zeta+a} & \dfrac{c}{\xi_{3}-\zeta+a}
\end{pmatrix}
\begin{pmatrix}
\gamma_{1}e^{{\rm i}A_{1}}\\
\gamma_{2}e^{{\rm i}A_{2}}\\
\gamma_{3}e^{{\rm i}A_{3}}
\end{pmatrix},
\end{equation}
where
\begin{align}
&D={\rm diag}\{1,e^{-{\rm i}\theta_{1}}, e^{-{\rm i}\theta_{2}}\},\label{21}\\
&A_{i}=\displaystyle\xi_{i}x-[A(t)+3\zeta B(t)]\xi_{i}^2+
[2\zeta A(t)+(a^2-4c^2+6\zeta^2)B(t)]\xi_{i}
+(4c^2+2\zeta^2)A(t)\nonumber\\&~~~~+(4c^2\zeta+2 a^2\zeta+6\zeta^3)B(t), \ i=1,2,3, \label{22}
\end{align}
with $\gamma_{i}$ being arbitrary real constants, and $\xi_{i}$ satisfying a cubic algebraic equation as
\begin{equation}\label{23}
\xi^3-(2c^2+a^2+3\zeta^2)\xi+2c^2\zeta-2 a^2\zeta+2\zeta^3=0.
\end{equation}
With the help of the basic solution (\ref{20}), one can work out different types
of explicit localized wave solutions,  or more specially  the rogue wave solutions
for Eqs. (\ref{01})-(\ref{02}). In the following, we present two different
categories of rogue wave solutions under different frequencies of the CW solution (\ref{03}).

\subsection{The case of  $a\neq \frac{1}{2}c$}

In this circumstance, it is found that the cubic algebraic equation (\ref{23}) does not have a
triple root. Nevertheless, by choosing adequate spectral parameter, one can
prove that there exists a complex double root for the algebraic equation.

Without loss of generality, we take $a=1,c=1$, and the spectral parameter be chosen as
\begin{equation}\label{24}
\zeta=\frac{z_{1}}{6}-\frac{\sqrt{3}}{2}\frac{{\rm i}}{z_{1}}+\bigg[-\dfrac{z_{1}}{12}+{\rm i}\bigg(\dfrac{z_{1}}{6}+\dfrac{\sqrt{3}}{12}z_{1}\bigg)\bigg]\epsilon^2,
\end{equation}
where
$z_{1}=\sqrt{6\sqrt{3}-9}$, and $\epsilon$ is a complex small perturbation parameter.
Then, we denote
\begin{align}
&\Phi_{1}=
D\begin{pmatrix}
1 & 1\\
\dfrac{1}{\xi_{1}-\zeta-1} & \dfrac{1}{\xi_{2}-\zeta-1}\\
\dfrac{1}{\xi_{1}-\zeta+1} & \dfrac{1}{\xi_{2}-\zeta+1}
\end{pmatrix}
\begin{pmatrix}
e^{{\rm i}A_{1}}\\
e^{{\rm i}A_{2}}
\end{pmatrix},\nonumber\\
&\Phi_{2}=
D\begin{pmatrix}
1 & 1\\
\dfrac{1}{\xi_{1}-\zeta-1} & \dfrac{1}{\xi_{2}-\zeta-1}\\
\dfrac{1}{\xi_{1}-\zeta+1} & \dfrac{1}{\xi_{2}-\zeta+1}
\end{pmatrix}
\begin{pmatrix}
-\dfrac{e^{{\rm i}A_{1}}}{\epsilon}\\
-\dfrac{e^{{\rm i}A_{2}}}{\epsilon}
\end{pmatrix},\nonumber
\end{align}
and
\begin{equation}
\Psi_{1}=(\psi_{1},\phi_{1},\chi_{1})^{T}=m\Phi_{1}
+n\Phi_{2}.
\end{equation}
Here $A_{i}$,  $\xi_{i}$ ($i=1,2$) are determined by Eqs. (\ref{22})-(\ref{23}) with $a=1,c=1$,
 $\zeta$ is fixed at (\ref{24}), and
\begin{align}
&m=m_{0}+m_{1}\epsilon^2+\cdots+m_{N-1}\epsilon^{2(N-1)},\nonumber\\
&n=n_{0}+n_{1}\epsilon^2+\cdots+n_{N-1}\epsilon^{2(N-1)},\nonumber
\end{align}
with
$m_{j}$, $n_{j}$ ($0\leq j\leq N-1$) being arbitrary real constants,
Until now, we can end up with a kind of $Nth$-order rogue wave solution for Eqs. (\ref{01})-(\ref{02}).

{\bf Proposition 2} The unified $Nth$-order rogue wave solution in the case of $a\neq \frac{1}{2}c$ (more precisely,
$a=1,c=1$) follows the compact form
\begin{align}
&u_{1}[N]=\sqrt{\dfrac{\alpha(t)}{\delta(t)}}e^{{\rm i}\theta_{1}}\dfrac{\det(M_{1})}{\det(M)},\label{s1}\\
&u_{2}[N]=\sqrt{\dfrac{\alpha(t)}{\delta(t)}}e^{{\rm i}\theta_{1}}\dfrac{\det(M_{2})}{\det(M)},\label{s2}
\end{align}
where
\begin{align}
&M=(M_{ij})_{1\leq i,j\leq N},\ M_{1}=M-3M[2]^{\dagger}M[1],\ M_{2}=M-3M[3]^{\dagger}M[1],\nonumber\\
&\theta_{1}=x+3A(t)-5B(t),\ \theta_{2}=-x+3A(t)+5B(t), \nonumber\\
&M_{ij}=\dfrac{1}{2(i-1)!2(j-1)!}\dfrac{\partial ^{2(i+j-2)}}{\partial \epsilon^{2(j-1)}\partial \epsilon^{*2(i-1)}}
\frac{\Psi_{1}^{\dagger}\Psi_{1}}{M_{d}}
\bigg|_{\epsilon\rightarrow0,\epsilon^{*}\rightarrow0},\nonumber\\
&M_{d}=-{\rm i}\dfrac{\sqrt{3}}{z_{1}}-\dfrac{z_{1}}{12}(\epsilon^2-\epsilon^{*2})
+{\rm i}\bigg(\dfrac{z_{1}}{6}+\dfrac{\sqrt{3}}{12}z_{1}\bigg)(\epsilon^2+\epsilon^{*2}),\nonumber\\
&M[1]=(\psi_{1}^{[0]},\psi_{1}^{[1]},\cdots, \psi_{1}^{[N-1]}),\
M[2]=(\phi_{1}^{[0]},\phi_{1}^{[1]},\cdots, \phi_{1}^{[N-1]})e^{{\rm i}\theta_{1}}, \nonumber\\
&M[3]=(\chi_{1}^{[0]},\chi_{1}^{[1]},\cdots, \chi_{1}^{[N-1]})e^{{\rm i}\theta_{2}},\
(\psi_{1}^{[l]},\phi_{1}^{[l]},\chi_{1}^{[l]})^{T}=\dfrac{1}{2l!}\dfrac{\partial^{2l}\Psi_{1}}
{\partial \epsilon^{2l}}\bigg|_{\epsilon\rightarrow0}. \nonumber
\end{align}

\subsection{The case of  $a=\frac{1}{2}c$}

In this situation, it is readily to verify that the cubic algebraic equation (\ref{23}) has a
triple root. As before, for concreteness, we set $a=\frac{1}{2}$, $c=1$  and the spectral parameter be
\begin{equation}\label{30}
\zeta=\dfrac{\sqrt{3}}{2}{\rm i}+\dfrac{\epsilon^3}{3},
\end{equation}
where $\epsilon$ is a complex small perturbation parameter as is mentioned before.
Afterwards, we introduce
\begin{align}
&\Phi_{1}=
D\begin{pmatrix}
1 & 1 & 1\\
\dfrac{1}{\xi_{1}-\zeta-\frac{1}{2}} & \dfrac{1}{\xi_{2}-\zeta-\frac{1}{2}} &\dfrac{1}{\xi_{3}-\zeta-\frac{1}{2}}\\
\dfrac{1}{\xi_{1}-\zeta+\frac{1}{2}} & \dfrac{1}{\xi_{2}-\zeta+\frac{1}{2}} &\dfrac{1}{\xi_{3}-\zeta+\frac{1}{2}}
\end{pmatrix}
\begin{pmatrix}
e^{{\rm i}A_{1}}\\
e^{{\rm i}A_{2}}\\
e^{{\rm i}A_{3}}
\end{pmatrix},\nonumber\\
&\Phi_{2}=
D\begin{pmatrix}
1 & 1 & 1\\
\dfrac{1}{\xi_{1}-\zeta-\frac{1}{2}} & \dfrac{1}{\xi_{2}-\zeta-\frac{1}{2}} &\dfrac{1}{\xi_{3}-\zeta-\frac{1}{2}}\\
\dfrac{1}{\xi_{1}-\zeta+\frac{1}{2}} & \dfrac{1}{\xi_{2}-\zeta+\frac{1}{2}} &\dfrac{1}{\xi_{3}-\zeta+\frac{1}{2}}
\end{pmatrix}
\begin{pmatrix}
\dfrac{e^{{\rm i}A_{1}}}{\epsilon}\\
\omega^{*}\dfrac{e^{{\rm i}A_{2}}}{\epsilon}\\
\omega\dfrac{e^{{\rm i}A_{3}}}{\epsilon}
\end{pmatrix},\nonumber\\
&\Phi_{3}=
D\begin{pmatrix}
1 & 1 & 1\\
\dfrac{1}{\xi_{1}-\zeta-\frac{1}{2}} & \dfrac{1}{\xi_{2}-\zeta-\frac{1}{2}} &\dfrac{1}{\xi_{3}-\zeta-\frac{1}{2}}\\
\dfrac{1}{\xi_{1}-\zeta+\frac{1}{2}} & \dfrac{1}{\xi_{2}-\zeta+\frac{1}{2}} &\dfrac{1}{\xi_{3}-\zeta+\frac{1}{2}}
\end{pmatrix}
\begin{pmatrix}
\dfrac{e^{{\rm i}A_{1}}}{\epsilon^2}\\
\omega\dfrac{e^{{\rm i}A_{2}}}{\epsilon^2}\\
\omega^{*}\dfrac{e^{{\rm i}A_{3}}}{\epsilon^2}
\end{pmatrix},\nonumber
\end{align}
and
\begin{equation}
\Psi_{1}=(\psi_{1},\phi_{1},\chi_{1})^{T}=m\Phi_{1}+n\Phi_{2}+s\Phi_{3}.
\end{equation}
Here $\omega=e^{2\pi{\rm i}/3}$,
$A_{i}$,  $\xi_{i}$ ($i=1,2,3$)
are determined by Eqs. (\ref{22})-(\ref{23}) with $a=\frac{1}{2},c=1$,
$\zeta$ is chosen as (\ref{30}), and
\begin{align}
&m=m_{0}+m_{1}\epsilon^3+\cdots+m_{N-1}\epsilon^{3(N-1)},\nonumber\\
&n=n_{0}+n_{1}\epsilon^3+\cdots+n_{N-1}\epsilon^{3(N-1)},\nonumber\\
&s=s_{0}+s_{1}\epsilon^3+\cdots+s_{N-1}\epsilon^{3(N-1)},\nonumber
\end{align}
with $m_{j}$, $n_{j}$ and
$s_{j}$ ($0\leq j\leq N-1$) being arbitrary real constants.
As a result, we can arrive at the other kind of $Nth$-order rogue wave solution for Eqs. (\ref{01})-(\ref{02}).

{\bf Proposition 3} The unified $Nth$-order rogue wave solution in the case of
$a= \frac{1}{2}c$ (or rather, $a=\frac{1}{2},c=1$) takes the simple determinant representation
\begin{align}
&u_{1}[N]=\sqrt{\dfrac{\alpha(t)}{\delta(t)}}e^{{\rm i}\theta_{1}}\dfrac{\det(S_{1})}{\det(S)},\label{s3}\\
&u_{2}[N]=\sqrt{\dfrac{\alpha(t)}{\delta(t)}}e^{{\rm i}\theta_{2}}\dfrac{\det(S_{2})}{\det(S)},\label{s4}
\end{align}
where
\begin{align}
&S=(S_{ij})_{1\leq i,j\leq N},\ S_{1}=S-3S[2]^{\dagger}S[1],\ S_{2}=S-3S[3]^{\dagger}S[1],\nonumber\\
&\theta_{1}=\dfrac{1}{2}x+\dfrac{15}{4}A(t)-\dfrac{23}{8}B(t),\ \theta_{2}=-\dfrac{1}{2}x+\dfrac{15}{4}A(t)+\dfrac{23}{8}B(t),\nonumber\\
&S_{ij}=\dfrac{1}{3(i-1)!3(j-1)!}\dfrac{\partial ^{3(i+j-2)}}{\partial \epsilon^{3(j-1)}\partial \epsilon^{*3(i-1)}}
\frac{\Psi_{1}^{\dagger}\Psi_{1}}{S_{d}}
\bigg|_{\epsilon\rightarrow0,\epsilon^{*}\rightarrow0},\nonumber\\
&S_{d}=\sqrt{3}{\rm i}+\dfrac{\epsilon^3}{3}-\dfrac{\epsilon^{*3}}{3},\nonumber\\
&S[1]=(\psi_{1}^{[0]},\psi_{1}^{[1]},\cdots, \psi_{1}^{[N-1]}),\
S[2]=(\phi_{1}^{[0]},\phi_{1}^{[1]},\cdots, \phi_{1}^{[N-1]})e^{{\rm i}\theta_{1}}, \nonumber\\
&S[3]=(\chi_{1}^{[0]},\chi_{1}^{[1]},\cdots, \chi_{1}^{[N-1]})e^{{\rm i}\theta_{2}},\
(\psi_{1}^{[l]},\phi_{1}^{[l]},\chi_{1}^{[l]})^{T}=\dfrac{1}{3l!}\dfrac{\partial^{3l}\Psi_{1}}
{\partial \epsilon^{3l}}\bigg|_{\epsilon\rightarrow0}. \nonumber
\end{align}

\section{Rogue waves with multiple compression points}

This section is devoted to study dynamics of the dark-bright and composite
rogue waves with multiple compression points provided by the propositions 2 and 3.

\subsection{Dark-bright rogue waves}

By resorting to Eqs. (\ref{s1})-(\ref{s2}) for $N=1$, one can obtain the rogue wave solution
with the form
\begin{align}
&u_{1}[1]=\sqrt{\dfrac{\alpha(t)}{\delta(t)}}e^{{\rm i}\theta_{1}}\dfrac{\rho_{1}(G_{1}+{\rm i}H_{1})}{F}, \label{s5}\\
&u_{2}[1]=\sqrt{\dfrac{\alpha(t)}{\delta(t)}}e^{{\rm i}\theta_{2}}\dfrac{\rho_{2}(G_{2}+{\rm i}H_{2})}{F},\label{s6}
\end{align}
where the corresponding polynomials and constants are given in appendix, and
we have set $\beta(t)=\nu \alpha(t)$  with $\nu$ being a small real constant and $m_{0}=5$, $m_{1}=0$.

To proceed,  as is discussed in \cite{30,35}, we suppose the GVD-TOD periodic modulation functions follow the form
\begin{equation}\label{alpha}
\alpha(t)=\beta(t)/\nu=1-d\cos(kt),
\end{equation}
and the SPM-XPM periodic modulation function be determined by the equation
\begin{equation}\label{delta}
\sqrt{\dfrac{\alpha(t)}{\delta(t)}}=1-\widetilde{d}\sin(\widetilde{k}t),
\end{equation}
where
$d$, $\widetilde{d}$ are the amplitudes of modulations, and $k$, $\widetilde{k}$ represent the frequencies.
At this point,  from Eq.  (\ref{AB}), we get
\begin{equation}
A(t)=t-\dfrac{d}{k}\sin(kt)+A_{0}.
\end{equation}

After that, we discuss the rogue wave dynamics described by Eqs. (\ref{s5})-(\ref{s6}).
As a matter of fact, it is readily to find that rogue wave solution in this case features the dark-bright structure, and
one can also compute that the maximum values of the wave amplitudes in $u_{1}$
and $u_{2}$ component both occur at $(x,A(t))=(2.11,-2.10)$. Hence, when
putting $A_{0}=-2.10$,
the number of the compression points for the rogue wave can be determined by the
number of the solutions to the following equation
\begin{equation}\label{ft}
t-\frac{d}{k}\sin(kt)=0.
\end{equation}
It is known that the explicit closed form expression for the solutions of the above transcendental equation is not
available. Whereas it is can be checked that as the value of $d$ increases, the number of the solutions increases in pairs.
To be specific, by using the compression points analysis \cite{35},
we denote
\begin{equation}
d_{\ell}=\dfrac{\pi}{2}+2\ell\pi,\ \ell\geq1,
\end{equation}
then
it is calculated that when $d$ increases from $d_{\ell}$ to $d_{\ell+1}$, the new solutions of Eq. (\ref{ft})
are generated and can be approximately localized at
$$t=\pm\dfrac{d_{\ell}}{k}.$$

Taking into account of the above facts, we can provide an exact calculation formula for the number of
the compression points, which is closely related to the value of the
amplitude of modulation.
We define it as $N_{cp}(d)$, then it holds that
\begin{equation}\label{NCP}
N_{cp}(d)=\left\{
\begin{array}{l}
1, d\in[0,1],\\
3, d\in(1,d_{1}),\\
4\ell+1, d=d_{\ell},\ \ell\geq1,\\
4\ell+3,d\in(d_{\ell},d_{\ell+1}),\ \ell\geq1.
\end{array}\right.
\end{equation}
In what follows, on basis of the formula (\ref{NCP}),
we present some concrete examples to illustrate the management of the
nonautonomous rogue waves with multiple compression points.

\begin{figure}[!h]
\centering
\renewcommand{\figurename}{{\bf Fig.}}
{\includegraphics[height=6cm,width=8.5cm]{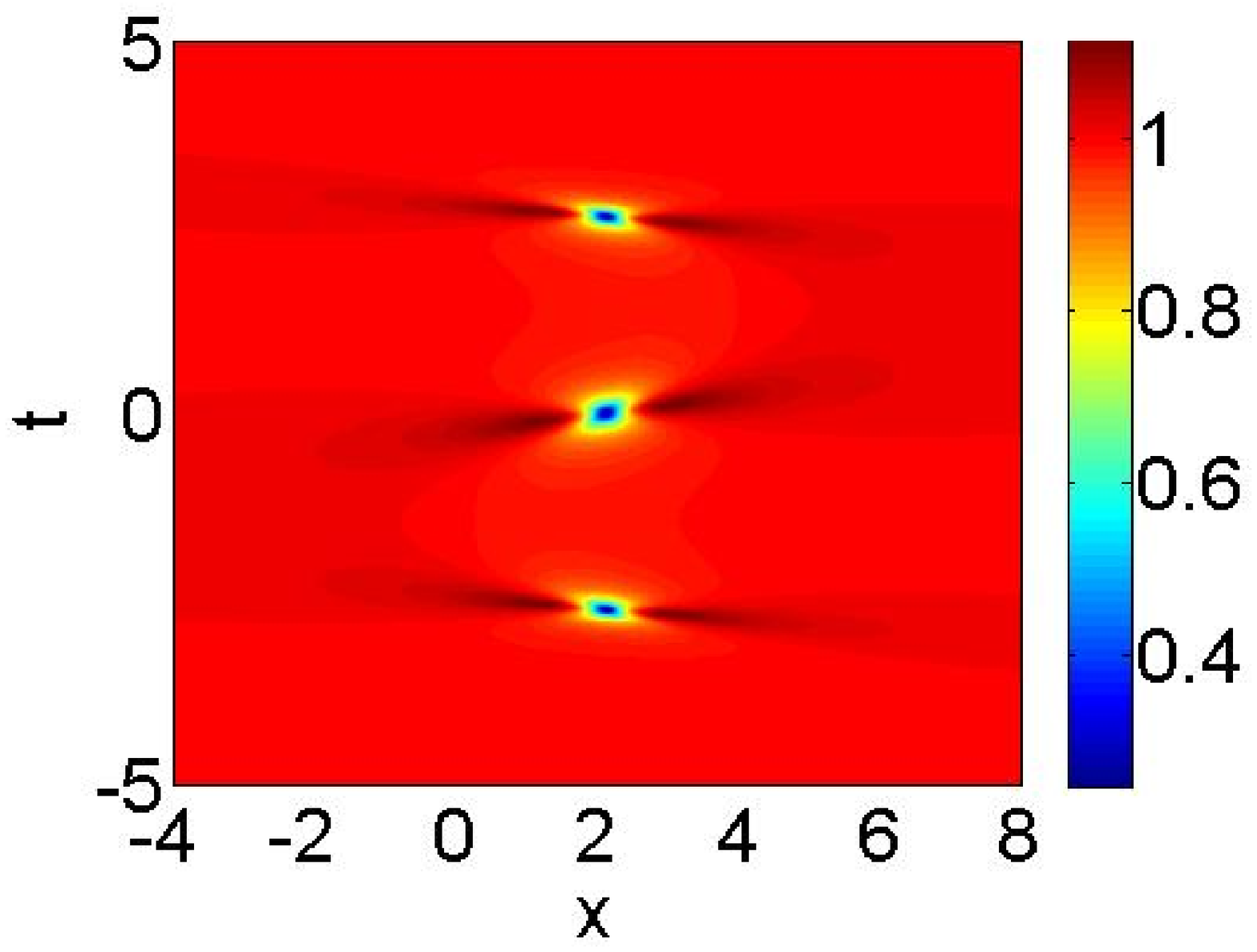}}
{\includegraphics[height=6cm,width=8.5cm]{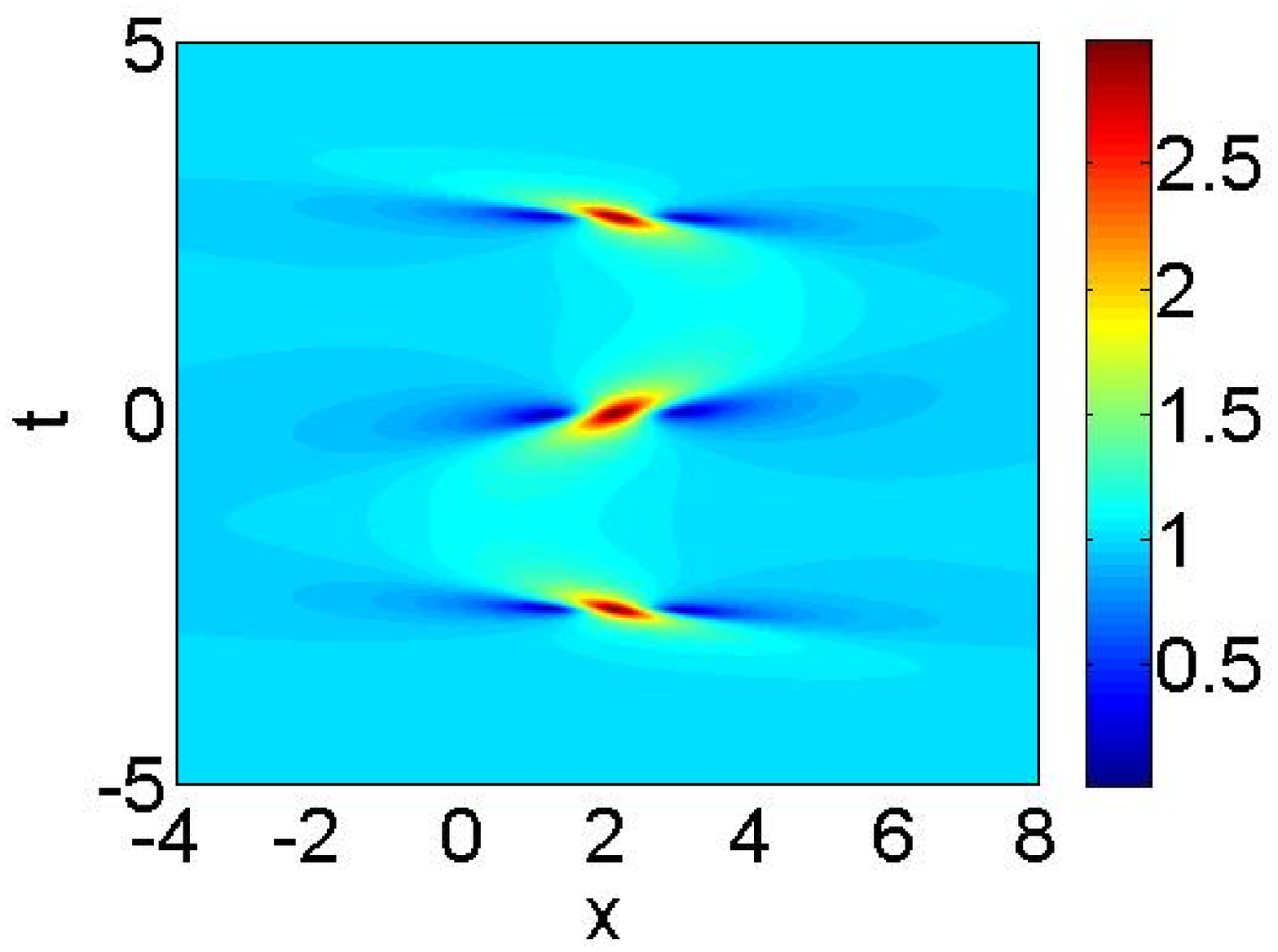}}
\begin{center}
\hskip 1cm $(\rm{a})$ \hskip 8cm $(\rm{b})$
\end{center}
\caption{Intensity of the solution (\ref{s5})-(\ref{s6}) with $d=3\pi/4,k=\pi/4, \widetilde{d}=0, A_{0}=-2.10$ and $\nu=0.1.$
(a), (b) Dark-bright rogue waves with triple compression points in $u_{1}$ and $u_{2}$ components.}
\label{fig:1}
\end{figure}

\begin{figure}[!h]
\centering
\renewcommand{\figurename}{{\bf Fig.}}
{\includegraphics[height=6cm,width=8.5cm]{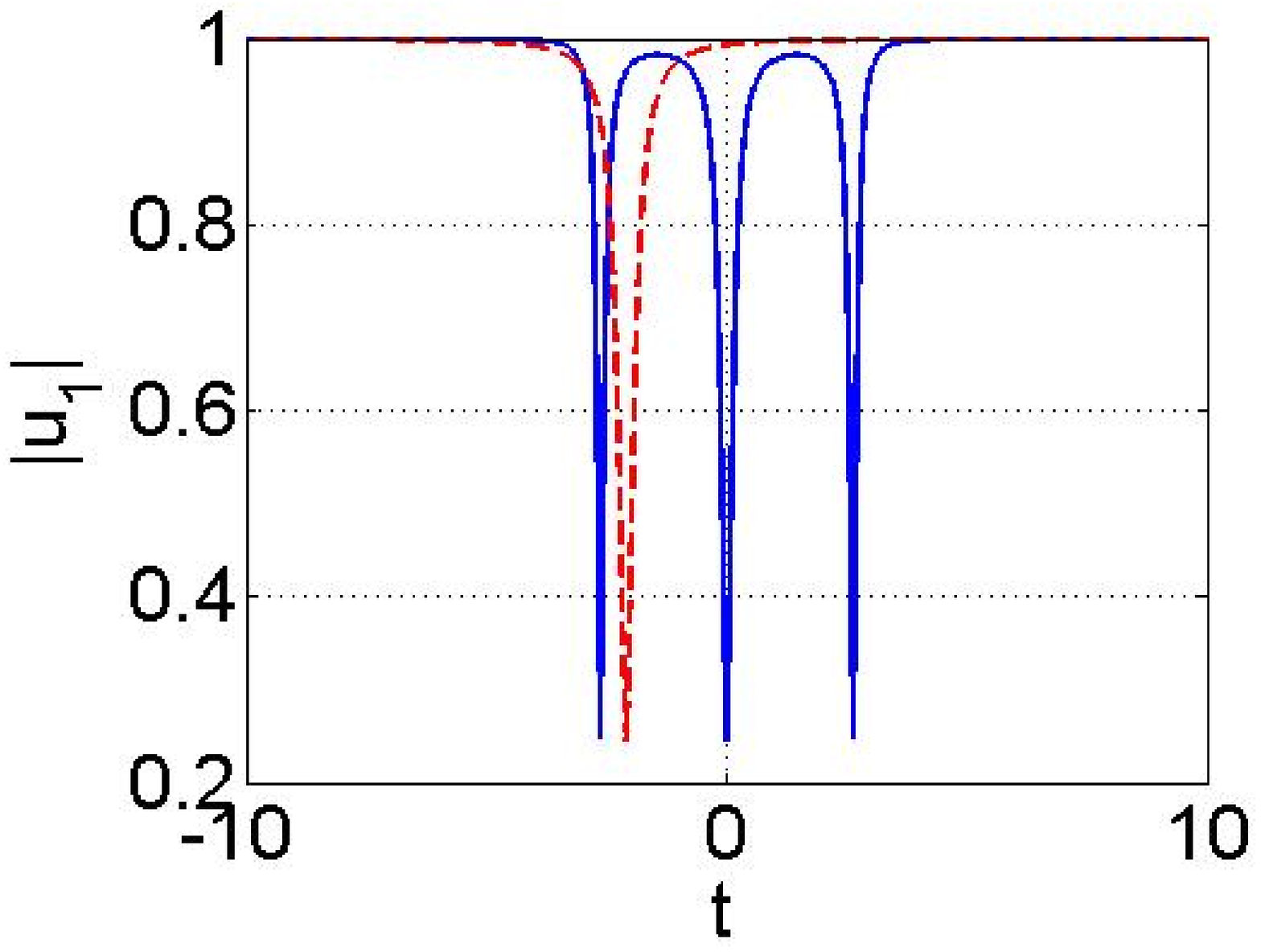}}
{\includegraphics[height=6cm,width=8.5cm]{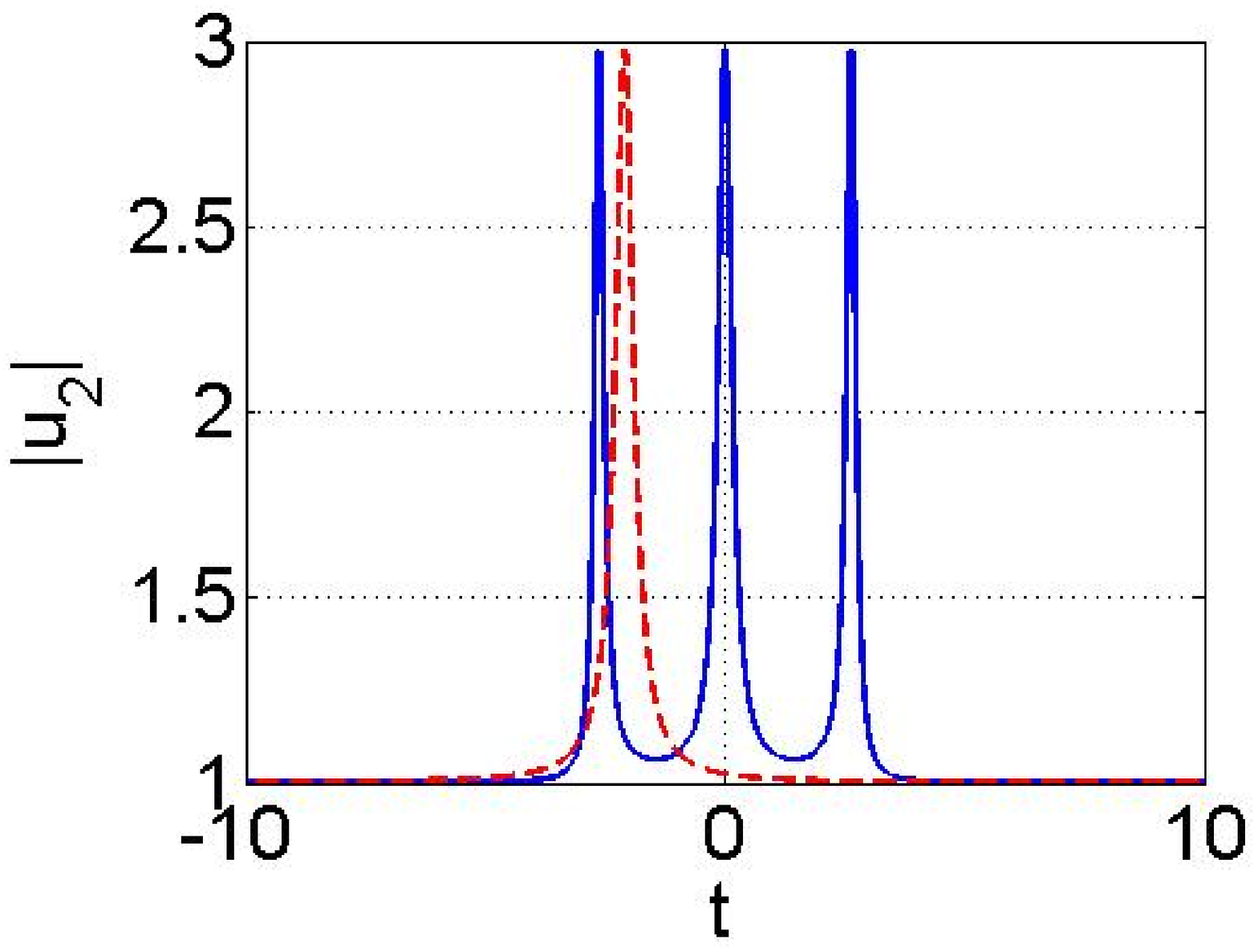}}
\begin{center}
\hskip 1cm $(\rm{a})$ \hskip 8cm $(\rm{b})$
\end{center}
\caption{Shape of the solution (\ref{s5})-(\ref{s6}) at $x=2.11$. (a), (b) Dark-bright
rogue waves with single (dashed line for $d=0$) and triple (solid line for $d=3\pi/4$) compression points.
The other parameters are the same as depicted in Fig. \ref{fig:1}. }
\label{fig:2}
\end{figure}

\begin{figure}[!h]
\centering
\renewcommand{\figurename}{{\bf Fig.}}
{\includegraphics[height=6cm,width=8.5cm]{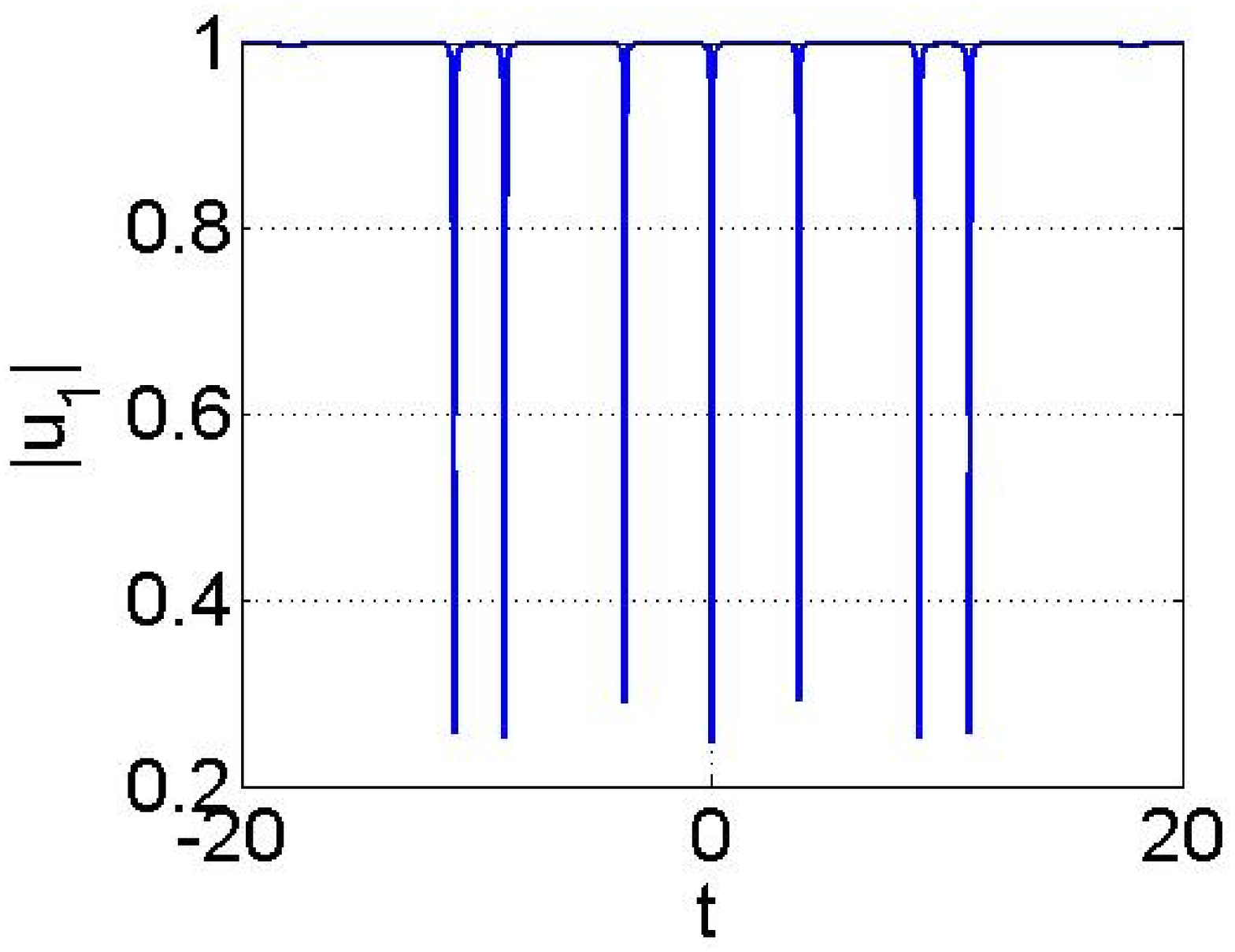}}
{\includegraphics[height=6cm,width=8.5cm]{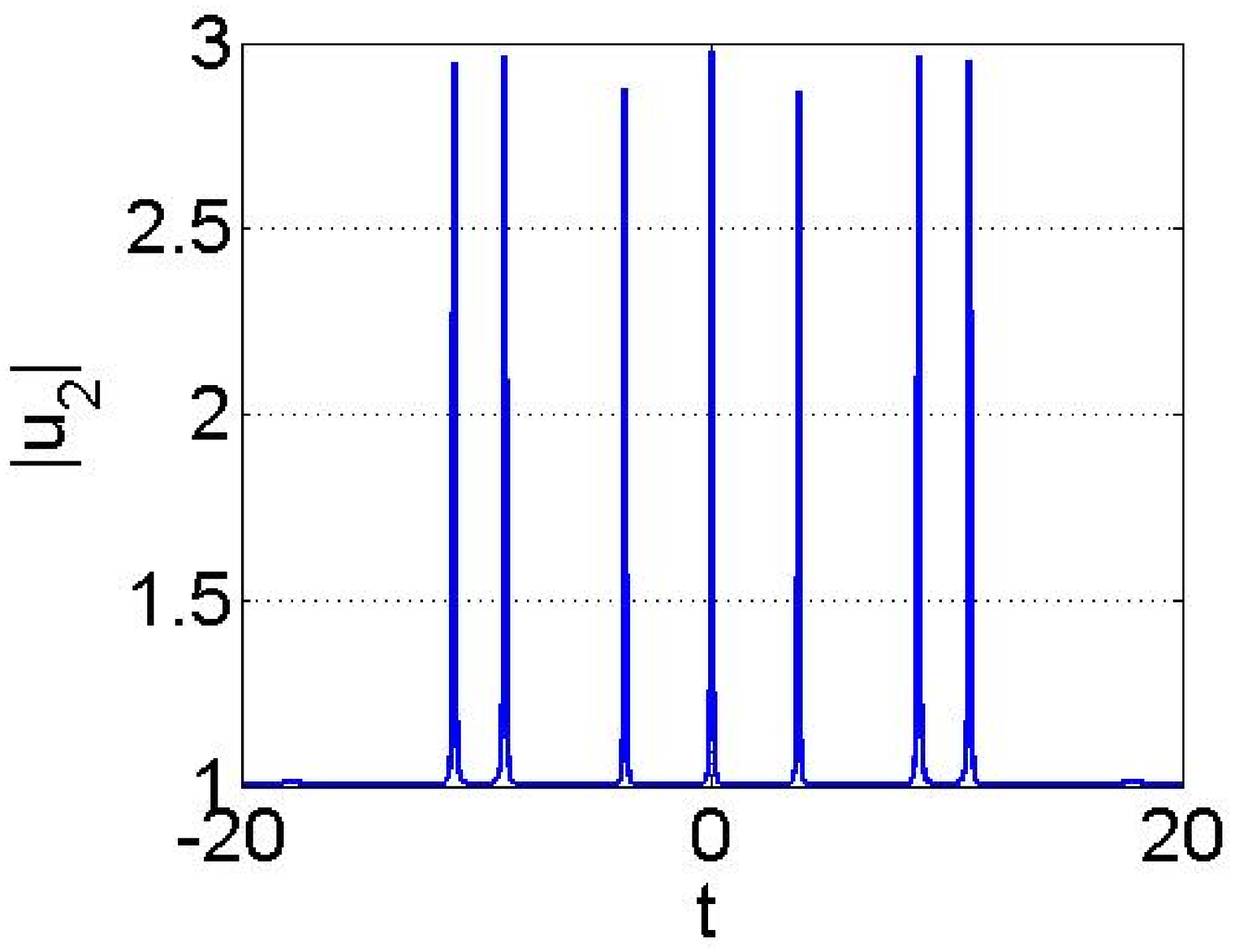}}
\begin{center}
\hskip 1cm $(\rm{a})$ \hskip 8cm $(\rm{b})$
\end{center}
\caption{(a), (b) Shape of dark-bright combs in $u_{1}$ and $u_{2}$ component with $d=15\pi/4$ at $x=2.11$.
The other parameters are the same as depicted in Fig. \ref{fig:1}.}
\label{fig:3}
\end{figure}

We first choose $d=3\pi/4\in(1,d_{1})$, then it is apparent to see that in Fig. \ref{fig:1},
there exist triple compression points for the dark-bright rogue waves, which can not be found in the
constant-coefficient systems.
While when taking $d\rightarrow0$ and $\widetilde{d}\rightarrow0$, the rogue wave solution (\ref{s5})-(\ref{s6}) is reduced into
the standard Peregrine soliton. The corresponding shape comparison between the standard Peregrine rogue wave
and the rogue wave with triple compression points is displayed in Fig. \ref{fig:2}. At this point,  it is  expressly to observe that
the single compression point split in two when $d$ from 0 to $3\pi/4$.
Moreover, as the value of $d$ is sufficiently large, for instance, by changing $d=15\pi/4\in(d_{1},d_{2})$,
then the dark-bright comb-like structures can be generated, see Fig. \ref{fig:3}.
Hereby, it is clearly seen that the dark-bright rogue waves
with sevenfold ($4\times1+3=7$) compression points come into being with a comb-like structure.
In addition, it is naturally to note that the frequency constant $k$ only controls the distance
among each multiple compression points, but it does not affect the main rogue wave characteristics.

\begin{figure}[!h]
\centering
\renewcommand{\figurename}{{\bf Fig.}}
{\includegraphics[height=6cm,width=8.5cm]{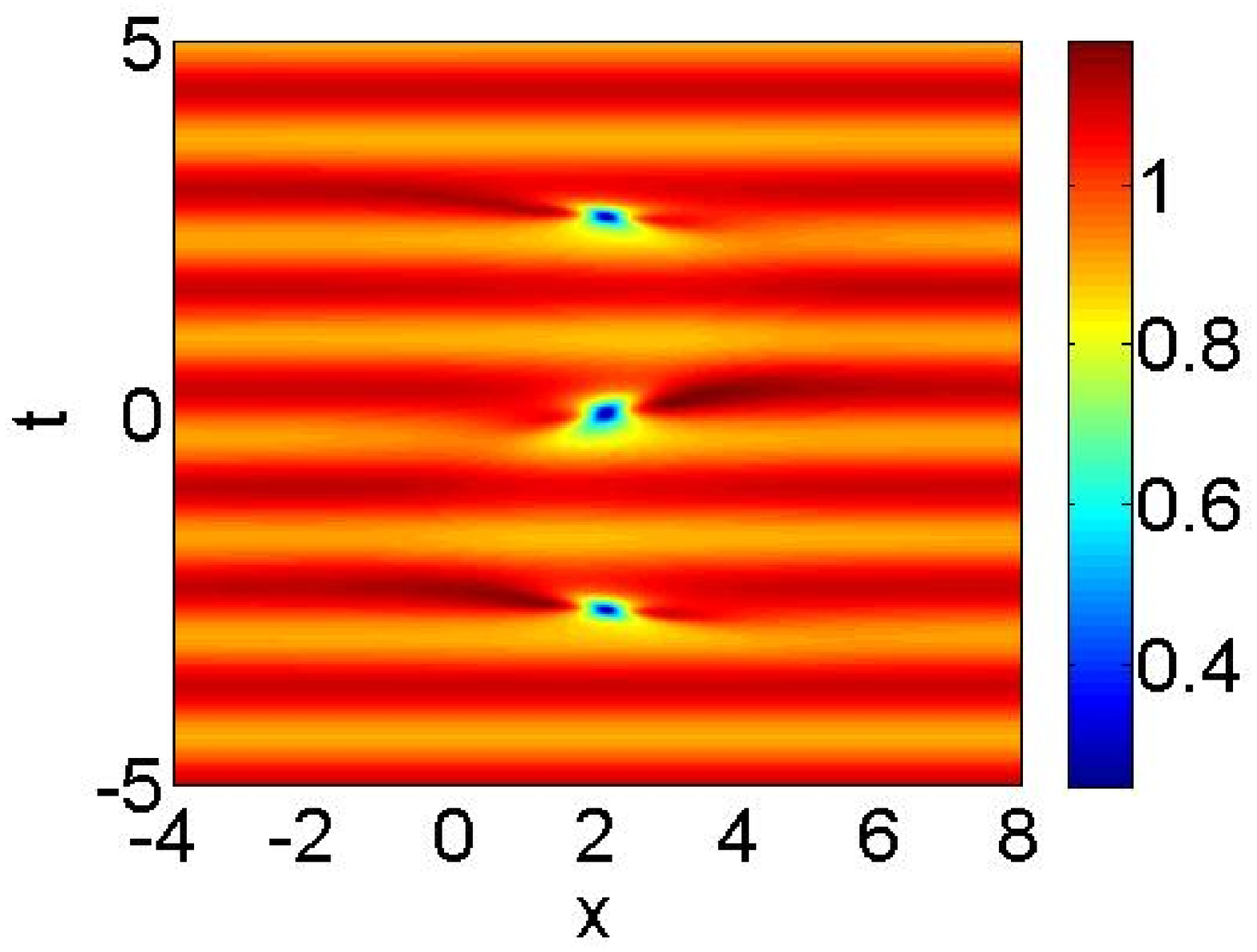}}
{\includegraphics[height=6cm,width=8.5cm]{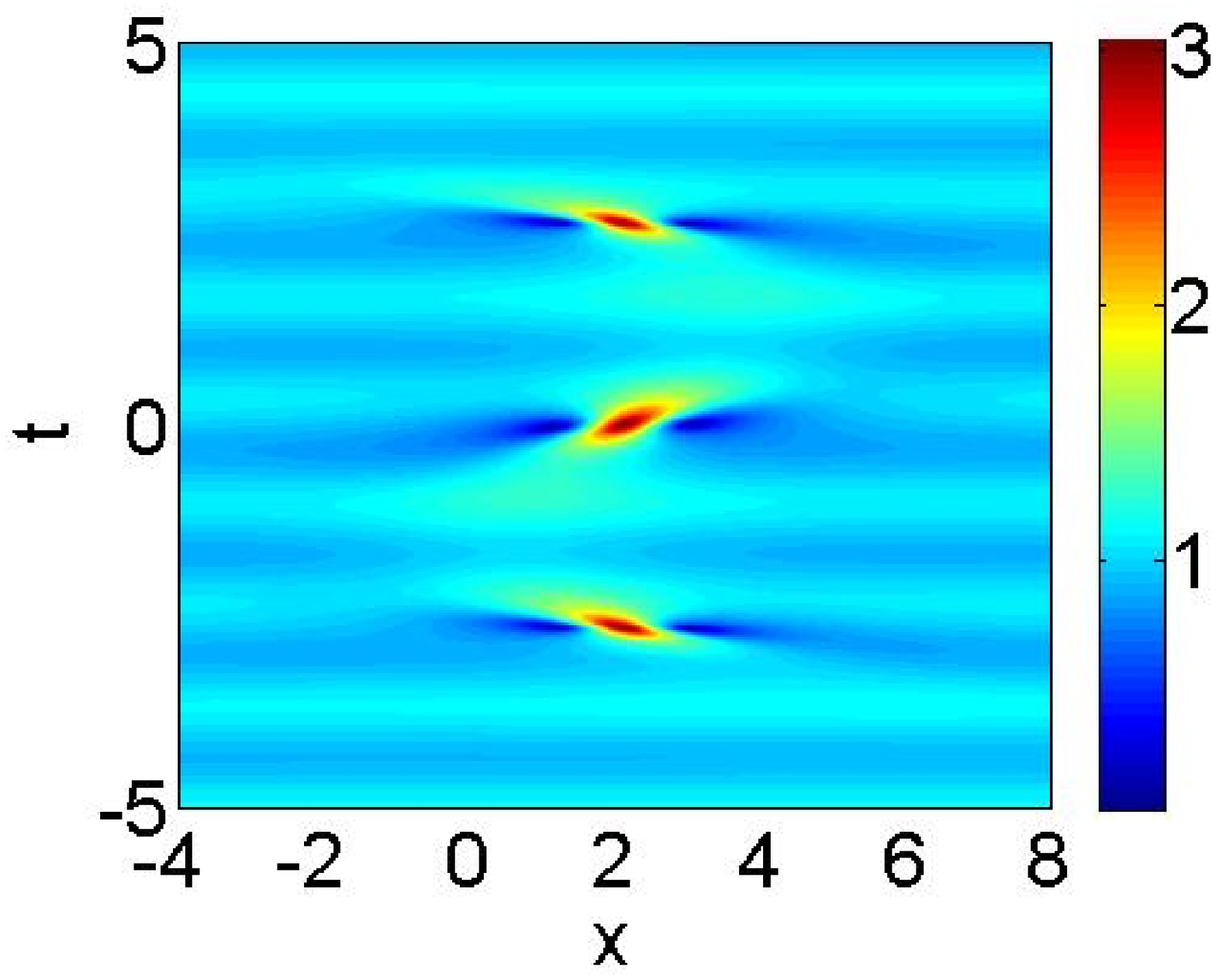}}
\begin{center}
\hskip 1cm $(\rm{a})$ \hskip 8cm $(\rm{b})$
\end{center}
\caption{Intensity of the solution (\ref{s5})-(\ref{s6}) with $d=3\pi/4,k=\pi/4, \widetilde{d}=-0.1, \widetilde{k}=3\pi/2,
A_{0}=-2.10$ and $\nu=0.1$. (a), (b) Dark-bright rogue waves with triple compression points under a sinusoidal wave
background in $u_{1}$ and $u_{2}$ components.}
\label{fig:4}
\end{figure}

Meanwhile, the propagation of rogue waves on periodic background can be
shown by controlling the values of the constants $\widetilde{d}$ and $\widetilde{k}$.
For the choice of specific parameters the intensity profiles of the dark-bright rogue waves with triple compression points
under the sinusoidal wave background are exhibited in Fig. \ref{fig:4}.
The constants $\widetilde{d}$ and $\widetilde{k}$ control the amplitude and
period of the background wave, respectively.

\begin{figure}[!h]
\centering
\renewcommand{\figurename}{{\bf Fig.}}
{\includegraphics[height=6cm,width=8.5cm]{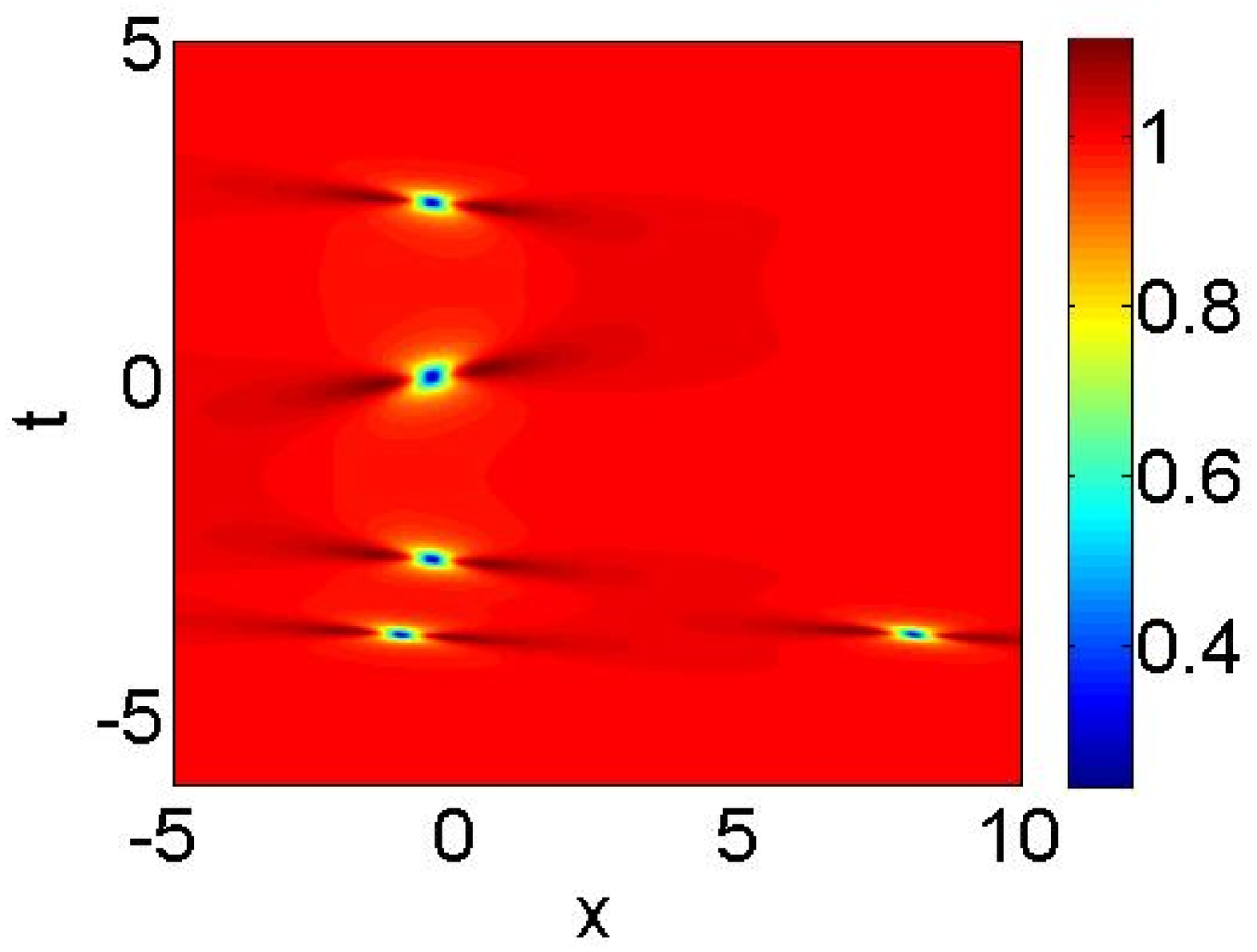}}
{\includegraphics[height=6cm,width=8.5cm]{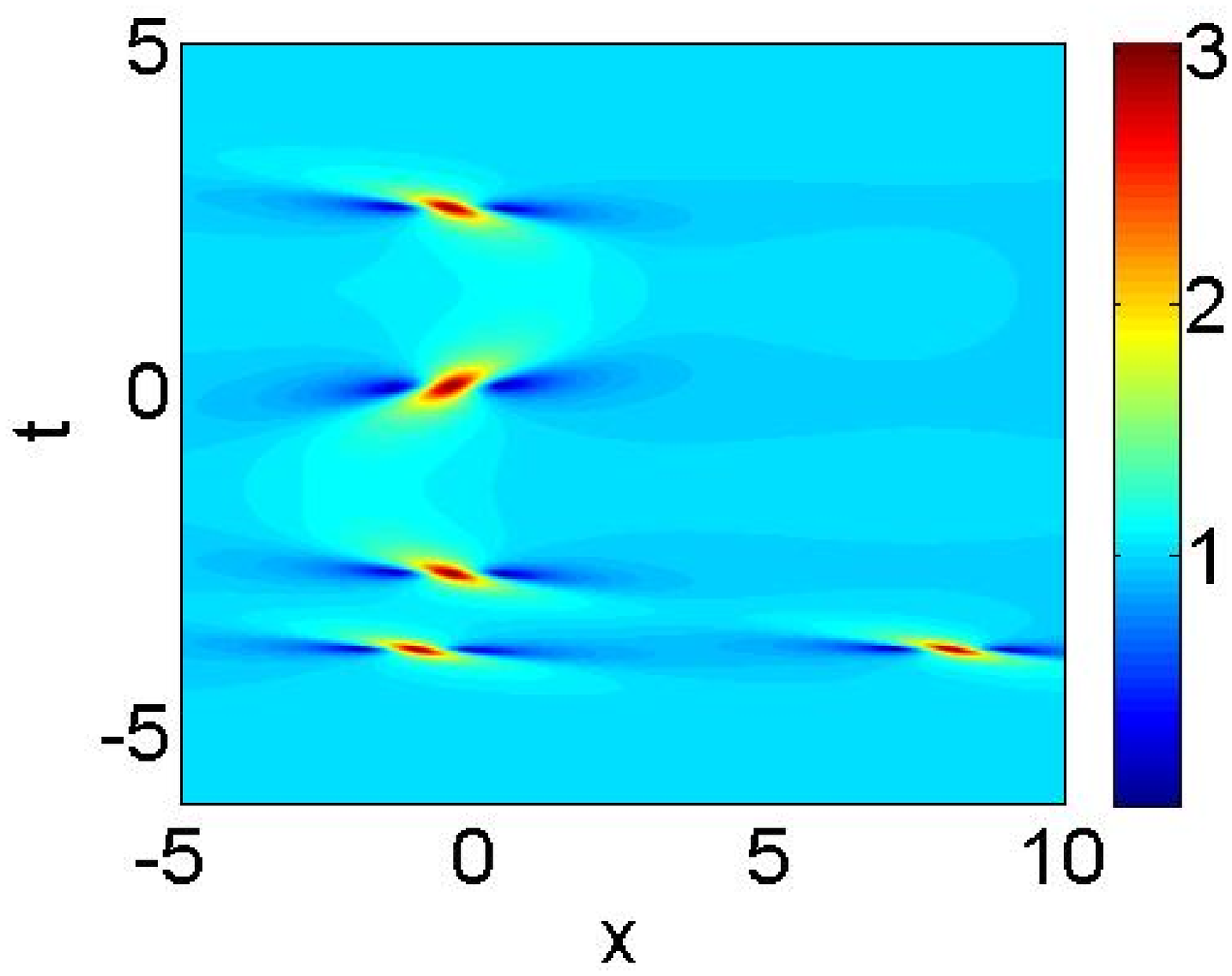}}
\begin{center}
\hskip 1cm $(\rm{a})$ \hskip 8cm $(\rm{b})$
\end{center}
\caption{Intensity of the second-order rogue wave solution given by Eqs.  (\ref{s1})-(\ref{s2})
with $N=2,d=3\pi/4,k=\pi/4, \widetilde{d}=0, m_{0}=5, m_{1}=0, n_{0}=1, n_{1}=0, A_{0}=0$ and  $\nu=0.1$.
(a), (b) Dark-bright three sisters involving one rogue wave with triple compression points
in $u_{1}$ and $u_{2}$ components. }
\label{fig:5}
\end{figure}

To proceed, by returning to Eqs. (\ref{s1})-(\ref{s2}) with $N=2$,
one can compute the explicit second-order rogue wave solution. Here we refrain from writing down
the cumbersome expression of the higher-order solution, and merely focus on demonstrating
the complex dynamics of it.
At this time, the second-order rogue waves can be classified into two types,
that is, the fundamental pattern and the triangular pattern (i.e. the celebrated \lq\lq three sisters" rogue wave) \cite{10}.
As before, in the light of the formula (\ref{NCP}), by varying the amplitude of the
GVD-TOD periodic modulation, the higher-order rogue waves with multiple compression points can emerge.
We show that in Fig. \ref{fig:5}(a), one dark rogue wave split into two extra peaks in the temporal direction at $t=0$
along with two normal dark rogue waves distribute with a triangle in $u_{1}$ component. Also, for the
bright case in $u_{2}$ component, the analogous wave phenomenon  occurs  as is seen in Fig. \ref{fig:5}(b).

\subsection{Composite rogue waves}

By means of the  formulas (\ref{s3})-(\ref{s4}) with $N=1$, we can develop
the rogue wave solution holding the form
\begin{align}
&u_{1}[1]=\sqrt{\dfrac{\alpha(t)}{\delta(t)}}e^{{\rm i}\theta_{1}}\dfrac{(\widetilde{G}_{1}+{\rm i}\widetilde{H}_{1})}{\widetilde{F}},\label{s7} \\
&u_{2}[1]=\sqrt{\dfrac{\alpha(t)}{\delta(t)}}e^{{\rm i}\theta_{2}}\dfrac{(\widetilde{G}_{2}+{\rm i}\widetilde{H}_{2})}{\widetilde{F}},\label{s8}
\end{align}
where the corresponding polynomials are presented in appendix,
and here we have taken $\beta(t)=\nu \alpha(t)$  with $\nu$ being a small real constant,
and $m_{0}=10$, $n_{0}=0$ and $s_{0}=1/10$.
At this time,  it is particular to point out that the above solution is characterized by
fourth degree polynomials. Consequently, there would appear composite states performing as rogue wave doublets (or rather pairs) when
plotting the intensity profiles of the rogue waves described by Eqs. (\ref{s7})-(\ref{s8}).

At this stage, as is mentioned in the above subsection, we identically select the periodic modulation functions
taking the forms of Eqs. (\ref{alpha})-(\ref{delta}). Further, through the
simple numerical calculations,
we can infer that the maximum values of the wave amplitude in $u_{1}$ component arrive at
$(x,A(t))=(14.13,0)$, $(x,A(t))=(-14.12,-0.08)$, and in $u_{2}$ component, the corresponding position coordinates are
$(x,A(t))=(14.13,0)$, $(x,A(t))=(-14.00,0.08)$.
In analogous to the dark-bright case, here we assume the integration constant be $A_{0}=0$.
Thus, with the help of the formula (\ref{NCP}), on can show the rogue wave doublets with multiple compression points.

\begin{figure}[!h]
\centering
\renewcommand{\figurename}{{\bf Fig.}}
{\includegraphics[height=6cm,width=8.5cm]{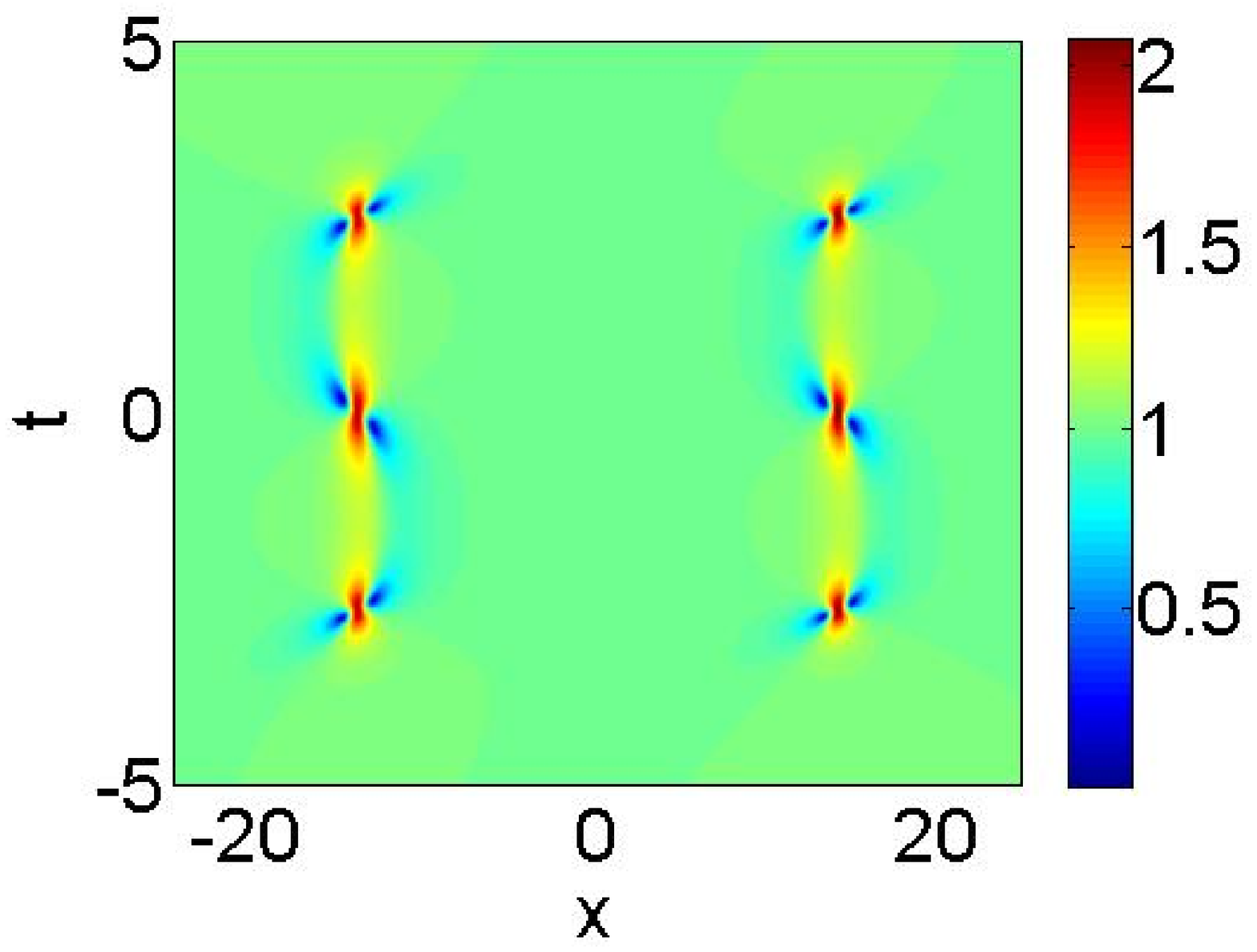}}
{\includegraphics[height=6cm,width=8.5cm]{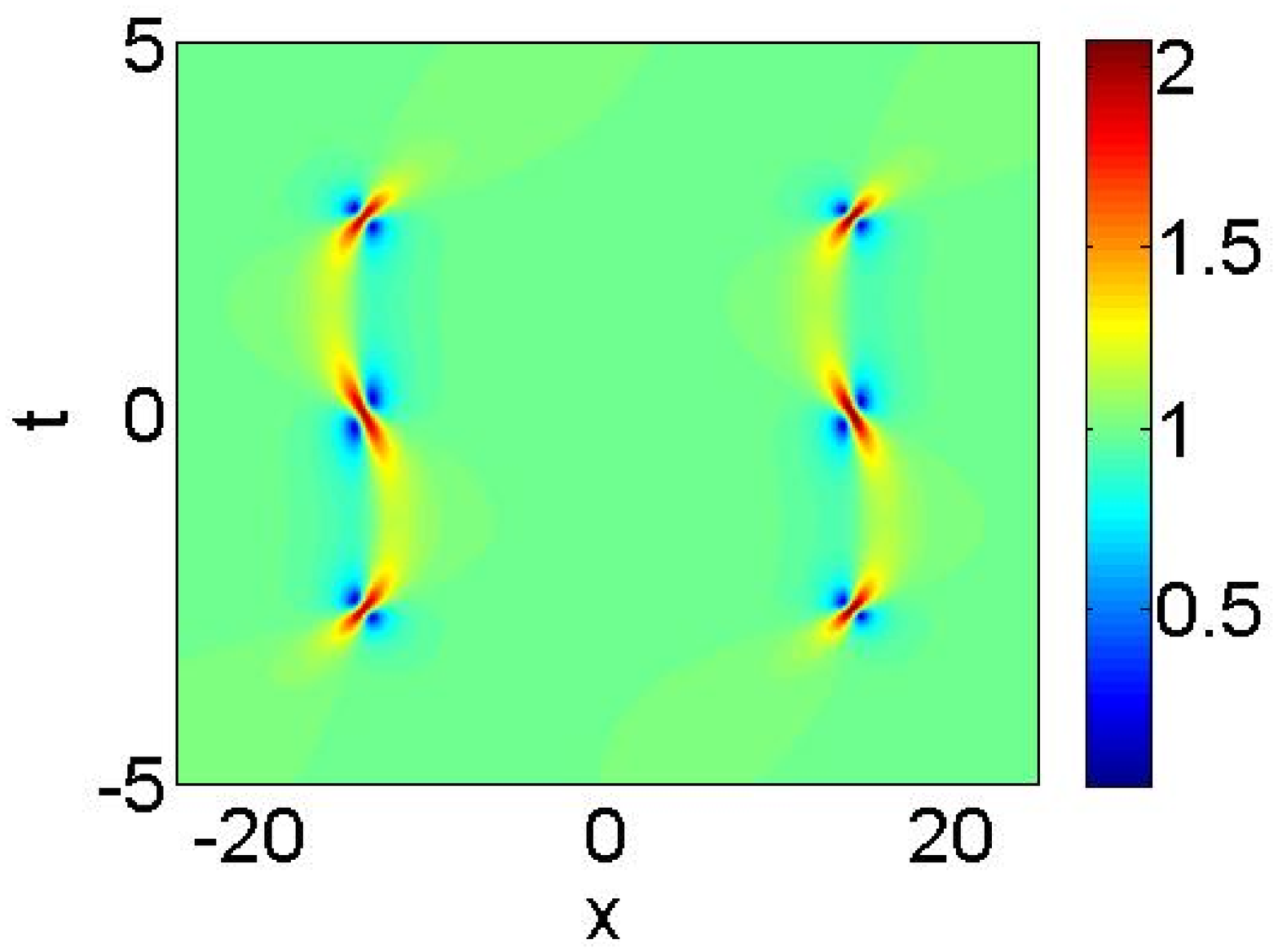}}
\begin{center}
\hskip 1cm $(\rm{a})$ \hskip 8cm $(\rm{b})$
\end{center}
\caption{Intensity of the solution (\ref{s7})-(\ref{s8}) with $d=3\pi/4,k=\pi/4, \widetilde{d}=0, A_{0}=0$ and $\nu=0.1.$
(a), (b) Rogue wave doublets with triple compression points in $u_{1}$ and $u_{2}$ components. }
\label{fig:6}
\end{figure}

\begin{figure}[!h]
\centering
\renewcommand{\figurename}{{\bf Fig.}}
{\includegraphics[height=6cm,width=8.5cm]{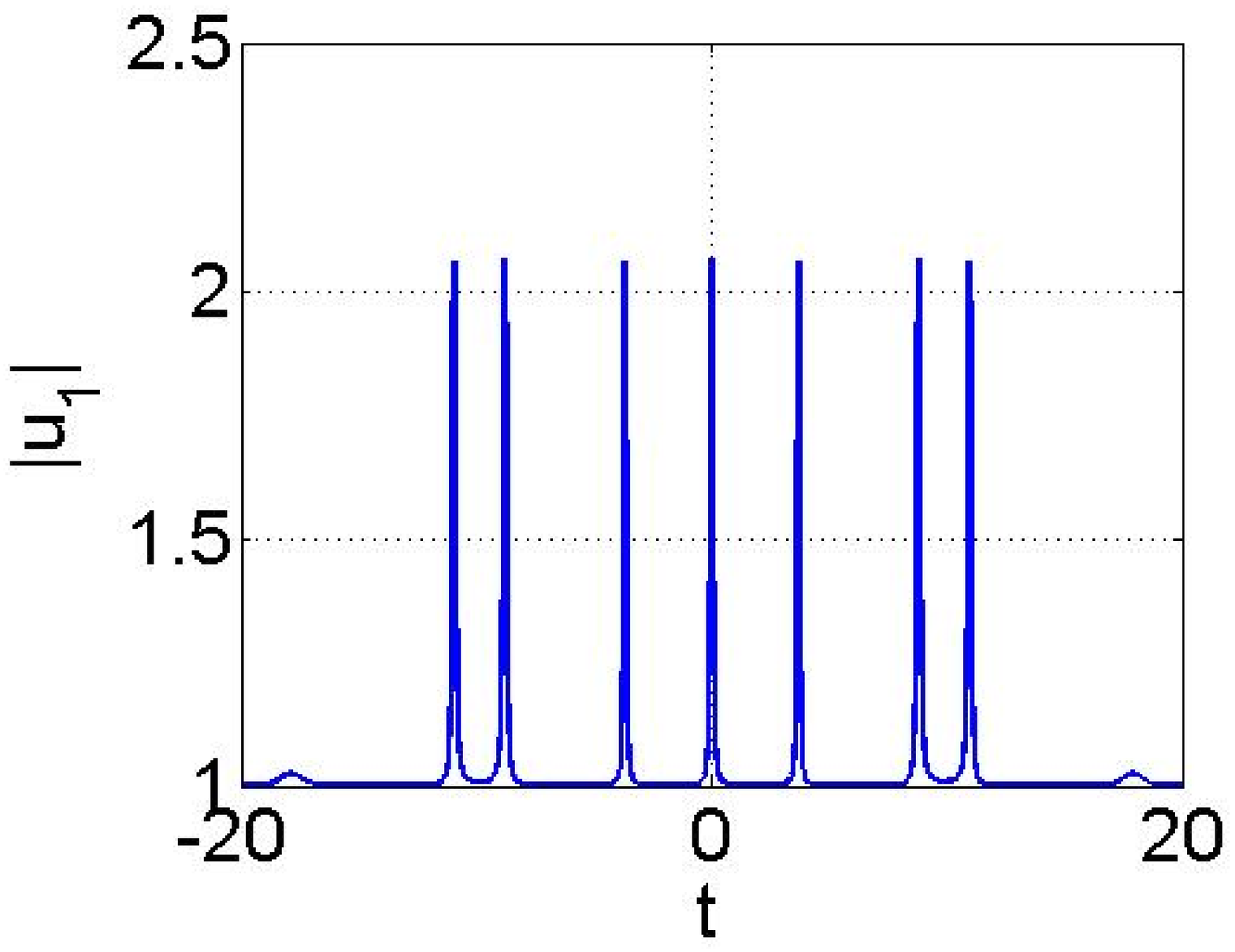}}
{\includegraphics[height=6cm,width=8.5cm]{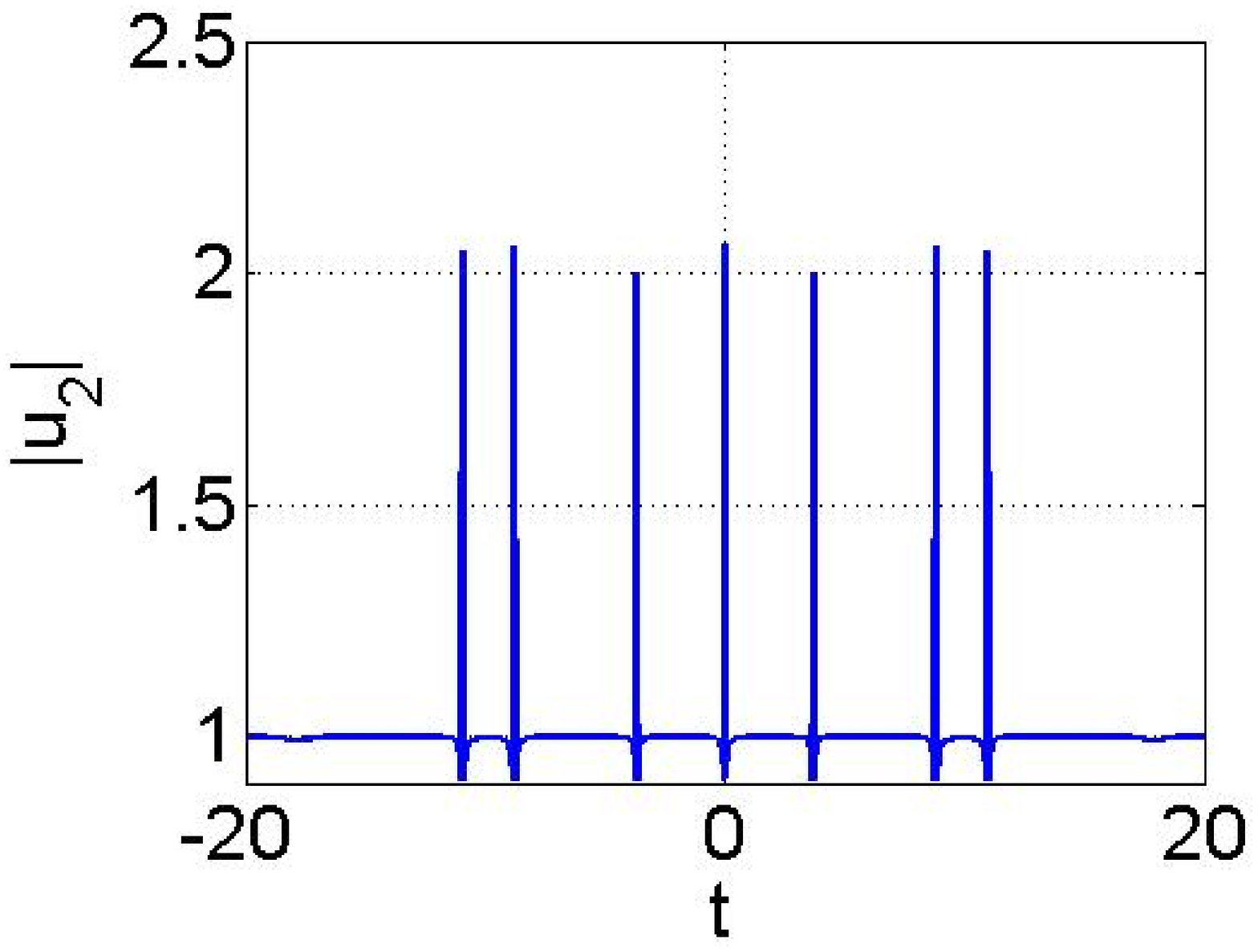}}
\begin{center}
\hskip 1cm $(\rm{a})$ \hskip 8cm $(\rm{b})$
\end{center}
\caption{(a), (b) Shape of comb doublets in $u_{1}$ and $u_{2}$ component with $d=15\pi/4$ at $x=14.13$.
The other parameters are the same as depicted in Fig. \ref{fig:6}. }
\label{fig:7}
\end{figure}

As is depicted in Figs. \ref{fig:6}, we observe that two rogue waves involving triple compression points
emerge on the temporal-spacial distribution plane in $u_{1}$ and $u_{2}$ components, respectively,
which is fairly different from the ordinary composite rogue waves in the coupled constant-coefficient system.
Meanwhile, in the similar way,  by sufficiently enlarging the value of
the amplitude of periodic modulation, the comb-like rogue wave doublets with sevenfold compression points can be produced, see Fig. \ref{fig:7}.
Here it is noteworthy that we only exhibit the shape of the comb-like structure at $x=14.13$, while for $x=-14.12$ or $x=-14.00$,
the situation is homologous yet the maximum values of the comb-like wave amplitudes are slightly
less than that of the actual sates, and we neglect this kind of difference in the present paper.

\begin{figure}[!h]
\centering
\renewcommand{\figurename}{{\bf Fig.}}
{\includegraphics[height=6cm,width=8.5cm]{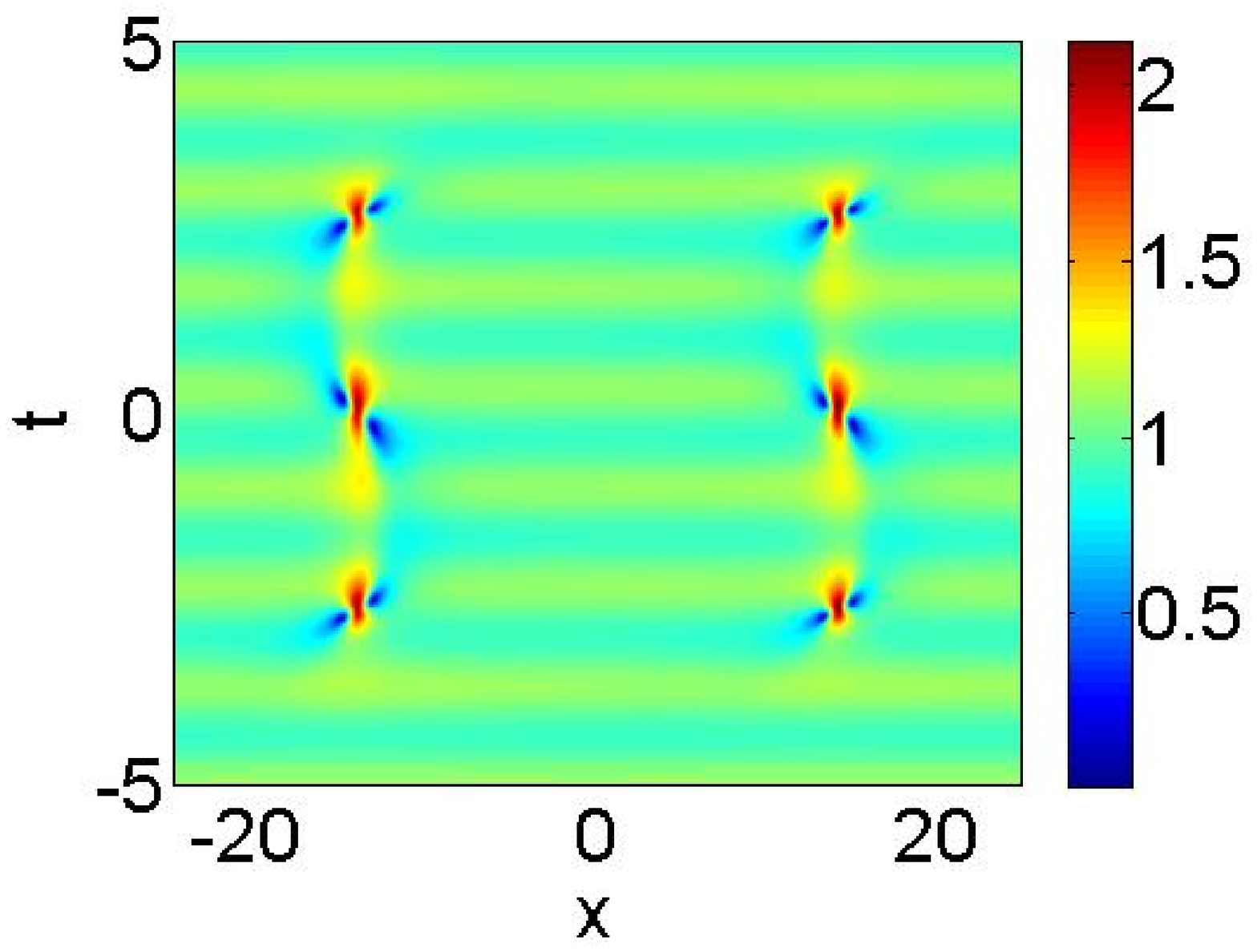}}
{\includegraphics[height=6cm,width=8.5cm]{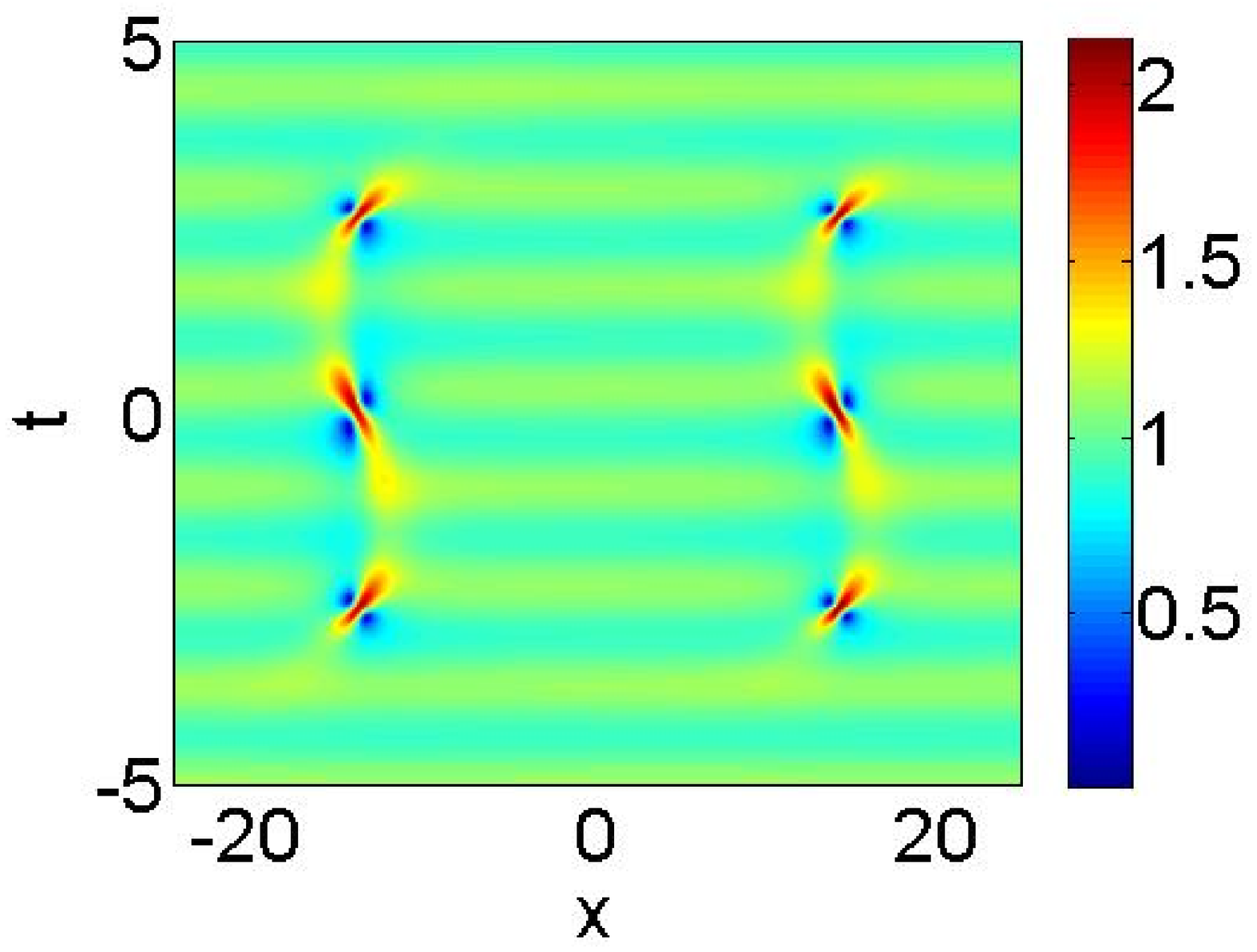}}
\begin{center}
\hskip 1cm $(\rm{a})$ \hskip 8cm $(\rm{b})$
\end{center}
\caption{Intensity of the solution (\ref{s7})-(\ref{s8}) with $d=3\pi/4,k=\pi/4, \widetilde{d}=-0.1, \widetilde{k}=3\pi/2,
A_{0}=0$ and $\nu=0.1$. (a), (b) Rogue wave doublets with triple compression points under a sinusoidal wave
background in $u_{1}$ and $u_{2}$ components.}
\label{fig:8}
\end{figure}

Besides, as is discussed before,
the rogue wave doublets with multiple compression points under periodic background can also be
presented, see Fig. \ref{fig:8}. The constants $\widetilde{d}$ and $\widetilde{k}$
play an identical part in influencing the SPM-XPM periodic modulation.

\begin{figure}[!h]
\centering
\renewcommand{\figurename}{{\bf Fig.}}
{\includegraphics[height=6cm,width=8.5cm]{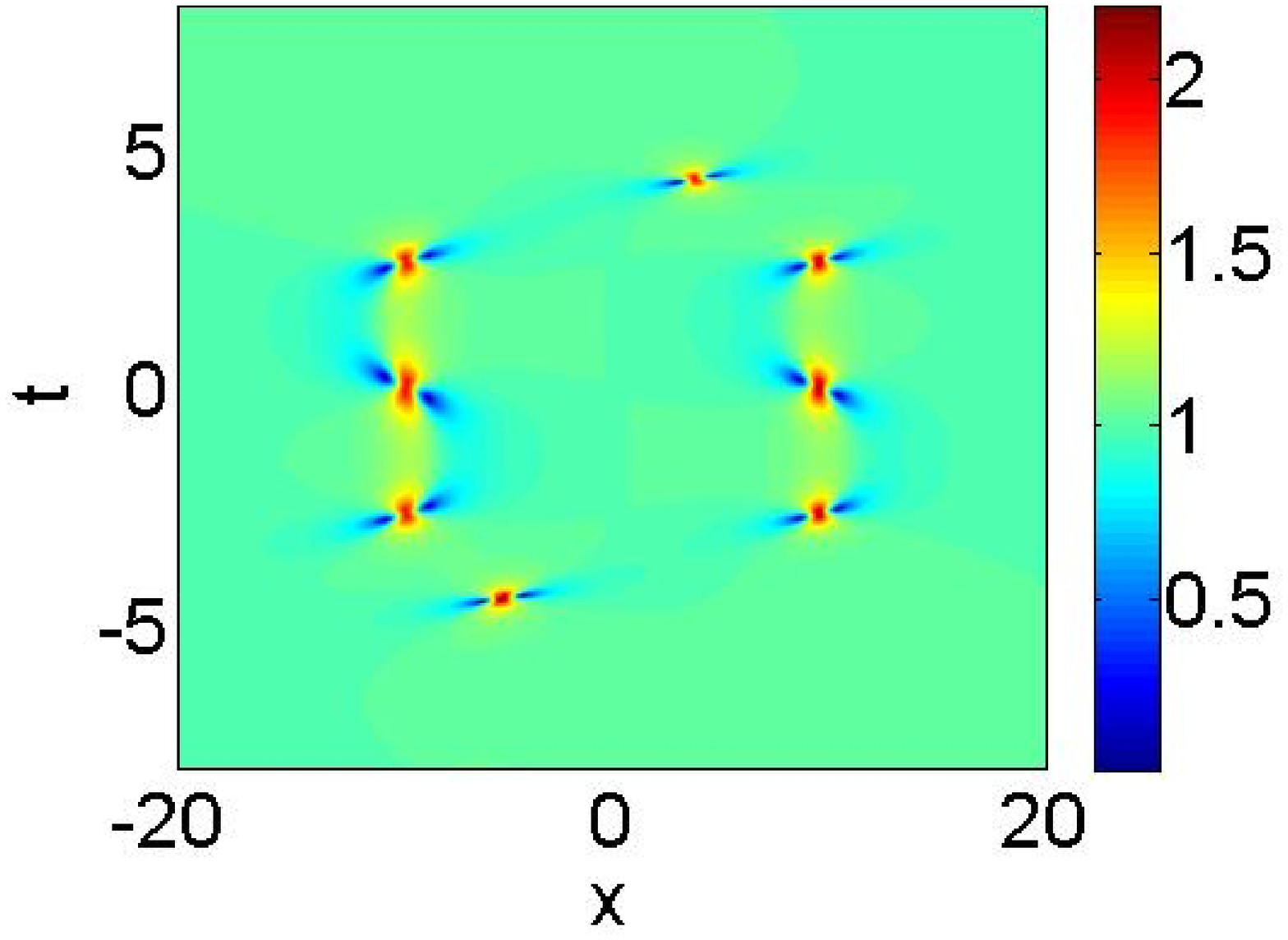}}
{\includegraphics[height=6cm,width=8.5cm]{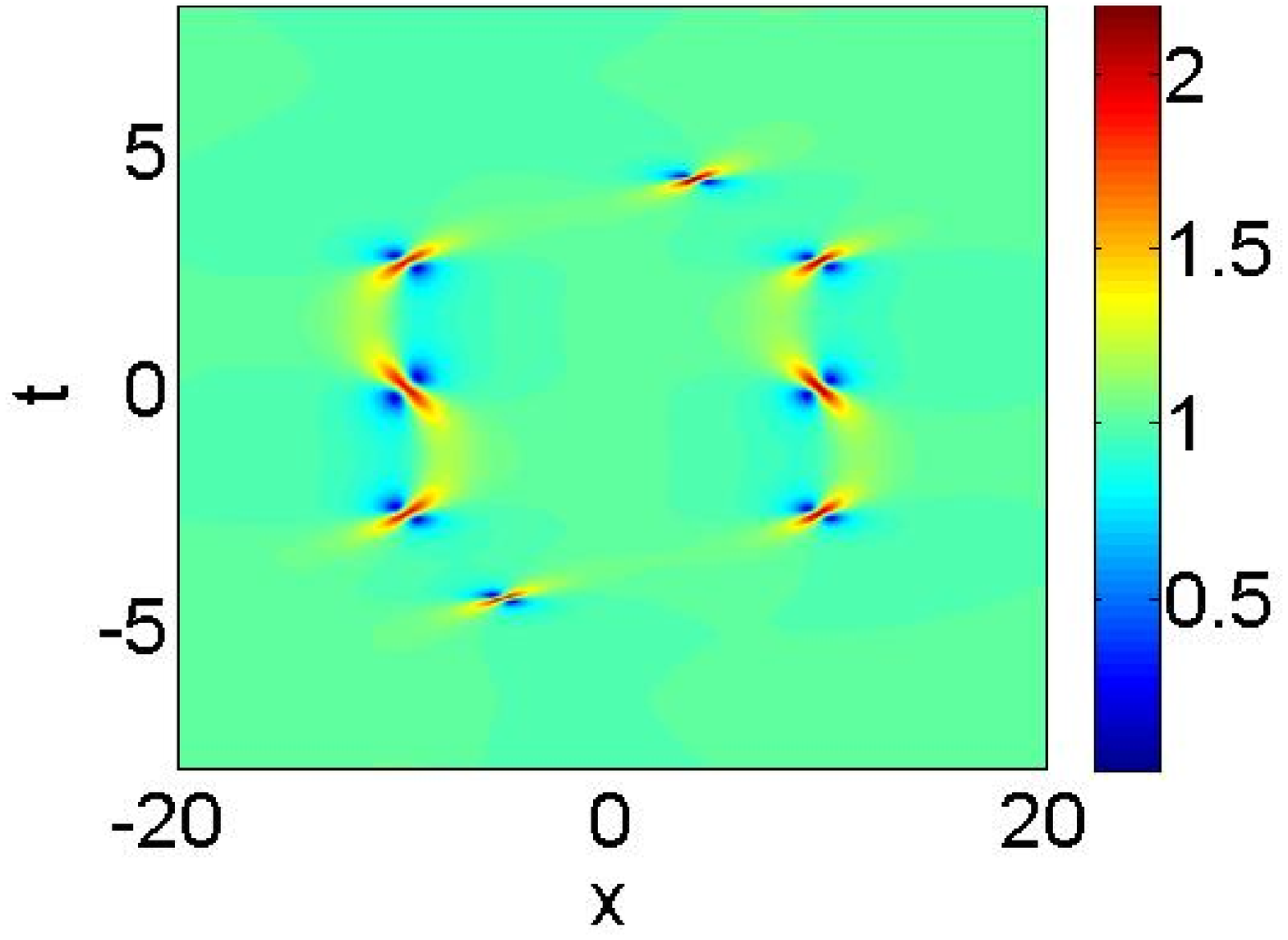}}
\begin{center}
\hskip 1cm $(\rm{a})$ \hskip 8cm $(\rm{b})$
\end{center}
\caption{Intensity of the second-order rogue wave solution given by Eqs. (\ref{s3})-(\ref{s4})
with $N=2,d=3\pi/4,k=\pi/4, \widetilde{d}=0, m_{0}=0, m_{1}=1000, n_{0}=1, n_{1}=0, s_{0}=0, s_{1}=0, A_{0}=0$ and $\nu=0.1$.
(a), (b) Rogue wave quartets involving two rogue waves with triple compression points
in $u_{1}$ and $u_{2}$ components. }
\label{fig:9}
\end{figure}

\begin{figure}[!h]
\centering
\renewcommand{\figurename}{{\bf Fig.}}
{\includegraphics[height=6cm,width=8.5cm]{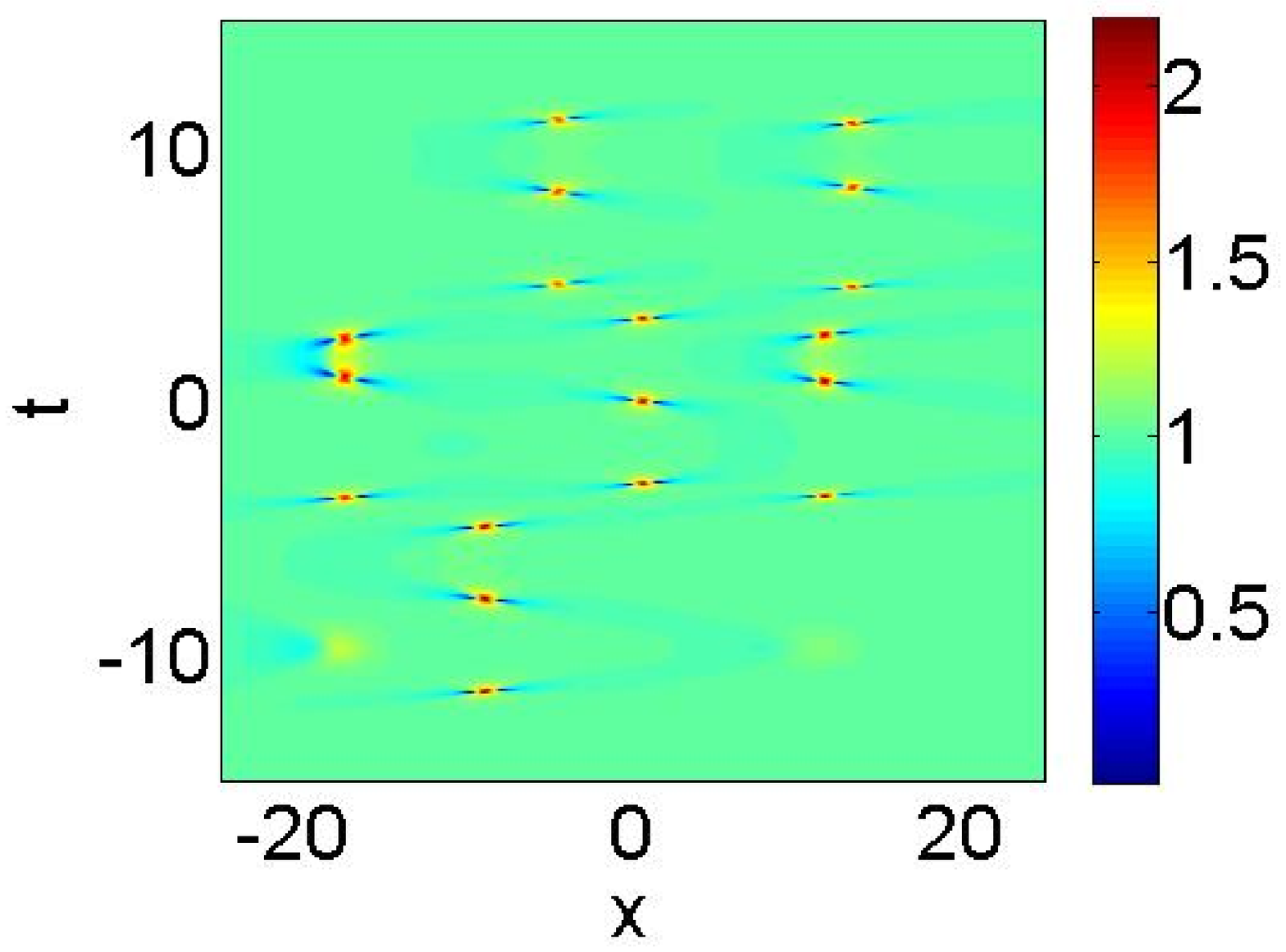}}
{\includegraphics[height=6cm,width=8.5cm]{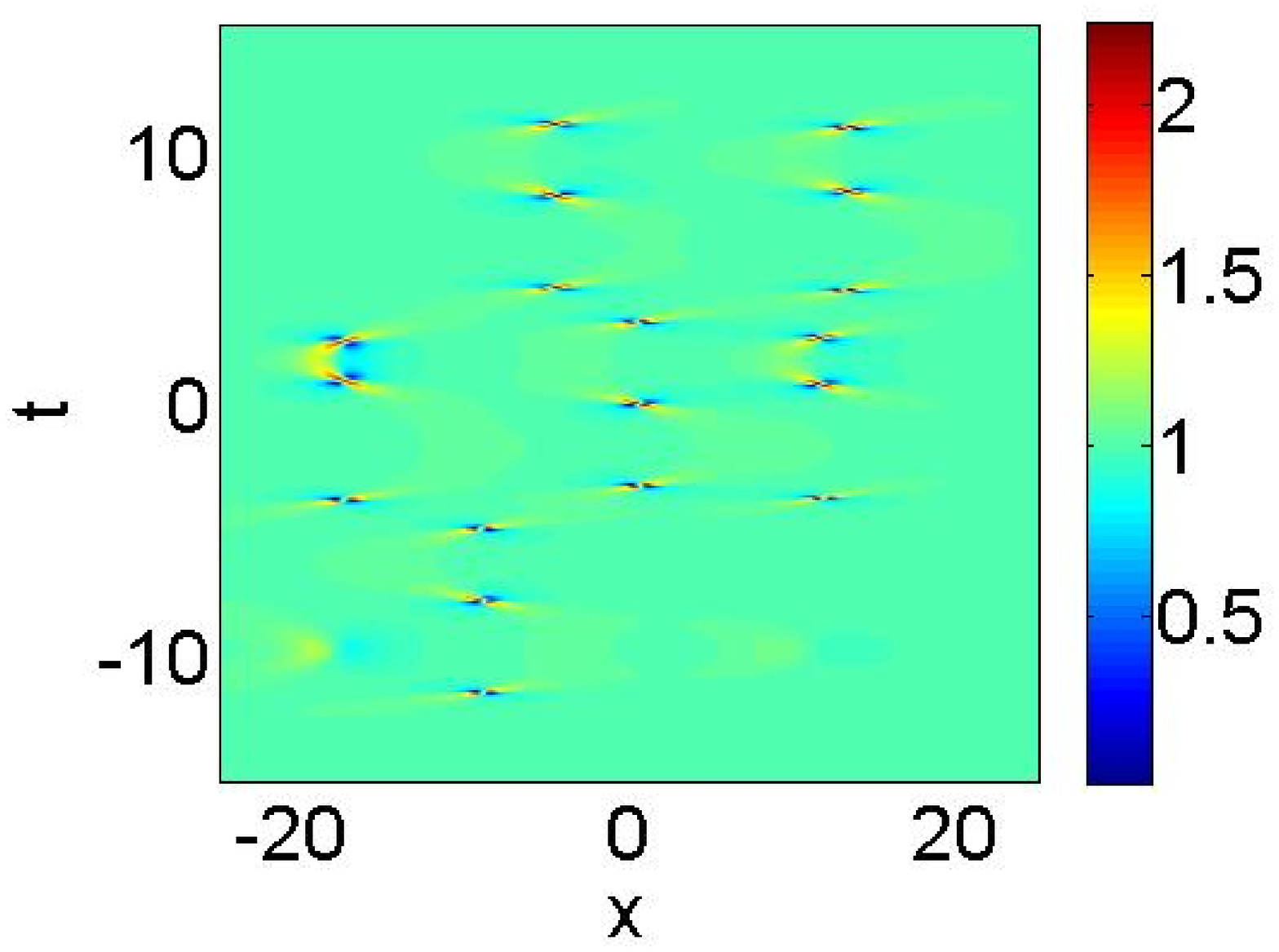}}
\begin{center}
\hskip 1cm $(\rm{a})$ \hskip 8cm $(\rm{b})$
\end{center}
\caption{Intensity of the second-order rogue wave solution given by Eqs.  (\ref{s3})-(\ref{s4})
with $N=2,d=5.8,k=\pi/4, \widetilde{d}=0, m_{0}=0, m_{1}=1000, n_{0}=0, n_{1}=0, s_{0}=0.1, s_{1}=0, A_{0}=0$ and $\nu=0.1$.
(a), (b) Rogue wave cluster involving six rogue waves with triple compression points
in $u_{1}$ and $u_{2}$ components.}
\label{fig:10}
\end{figure}

Ulteriorly, one can carry out the second-order rogue wave solution by means of Eqs. (\ref{s3})-(\ref{s4})
with $N=2$. In this status, the solution is constituted by the ratio of eighth or twelfth-order
polynomials. As a consequence, the higher-order rogue waves are
composite of four or six fundamental rogue waves and would exhibit more abundant patterns.
Naturally, for the variable-coefficient case, the corresponding
rogue wave quartets or sextets involving one or more rogue waves with multiple compression points (namely, the rogue wave
cluster) can arise by adjusting the values of the periodic modulation coefficients.
As an example, it is vividly displayed that in Fig.  \ref{fig:9}, two rogue waves in the direction of $t=0$
separate into three compression points, together with two normal rogue waves appear with a quadrilateral
in $u_{1}$ and $u_{2}$ components, respectively. In Fig. \ref{fig:10}, we see that
the rogue wave cluster including six rogue waves with triple compression points are shown.
It is notable to remark that although the nonautonomous rogue wave quartets or sextets depicted in Figs. \ref{fig:9}-\ref{fig:10}
in $u_{1}$ and $u_{2}$ components are similar, yet the concrete maximum wave amplitudes and their coordinate positions
are not identical in each component.

In the last of this section, we would like to mention that the rogue wave solutions given in this paper can be
reduced to the Manakov system by taking $\nu\rightarrow 0$ and $\delta(t)/\alpha(t)$ tend to a constant.
On account of the presence of the higher-order effects, one can find that
under identical parameter conditions compared to the Manakov system \cite{20,21,22,23},
the \lq ridge\rq \ \cite{15,26} of the rogue waves in Eqs. (\ref{01})-(\ref{02}) is tilted to a certain extent,
and the distributions of the rogue waves in t dimension is contracted significantly, see the intensity
profiles in Figs. \ref{fig:1}-\ref{fig:10}.

\section{Further properties of the dark rogue wave solution }

\begin{figure}[!h]
\centering
\renewcommand{\figurename}{{\bf Fig.}}
{\includegraphics[height=6cm,width=8.5cm]{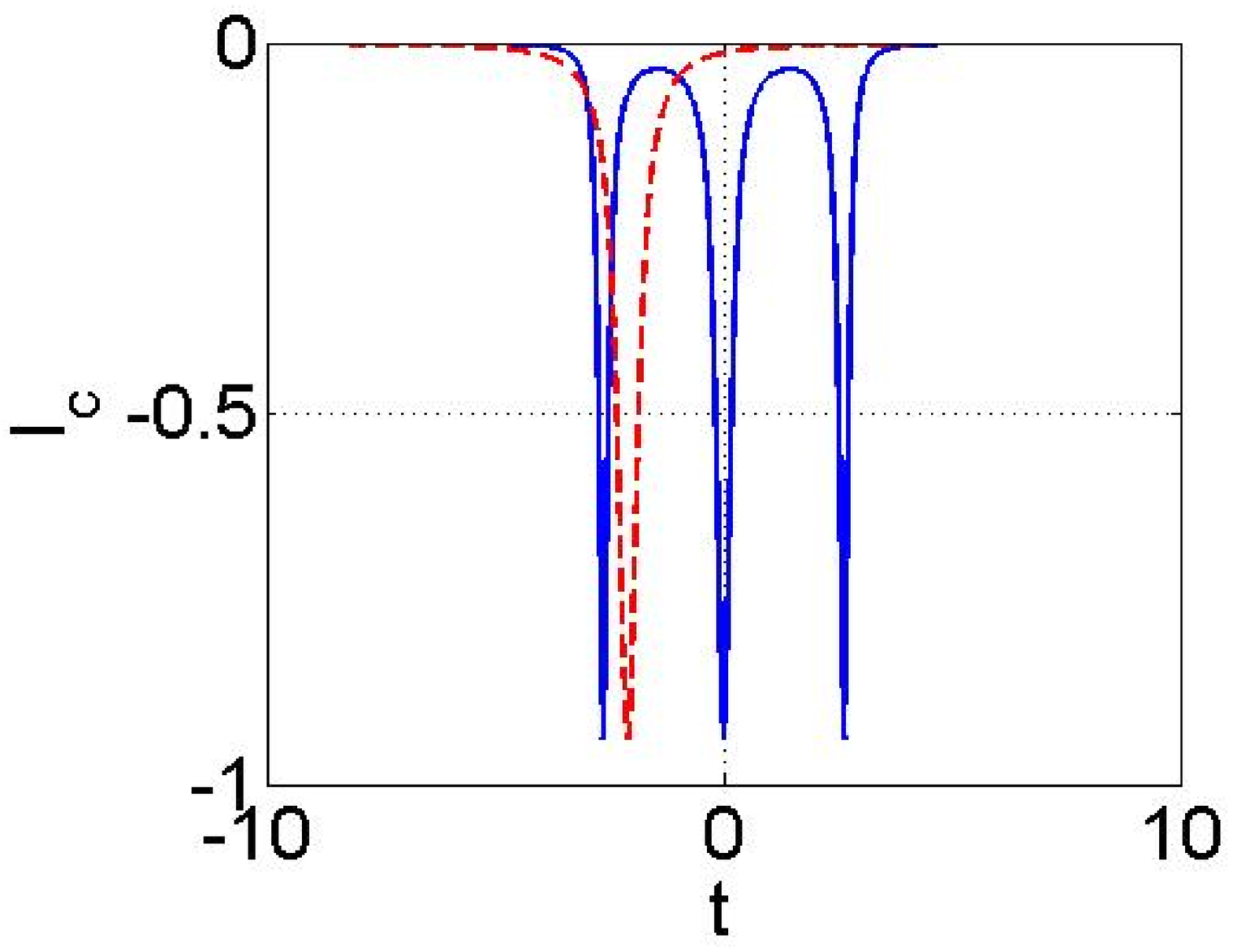}}
{\includegraphics[height=6cm,width=8.5cm]{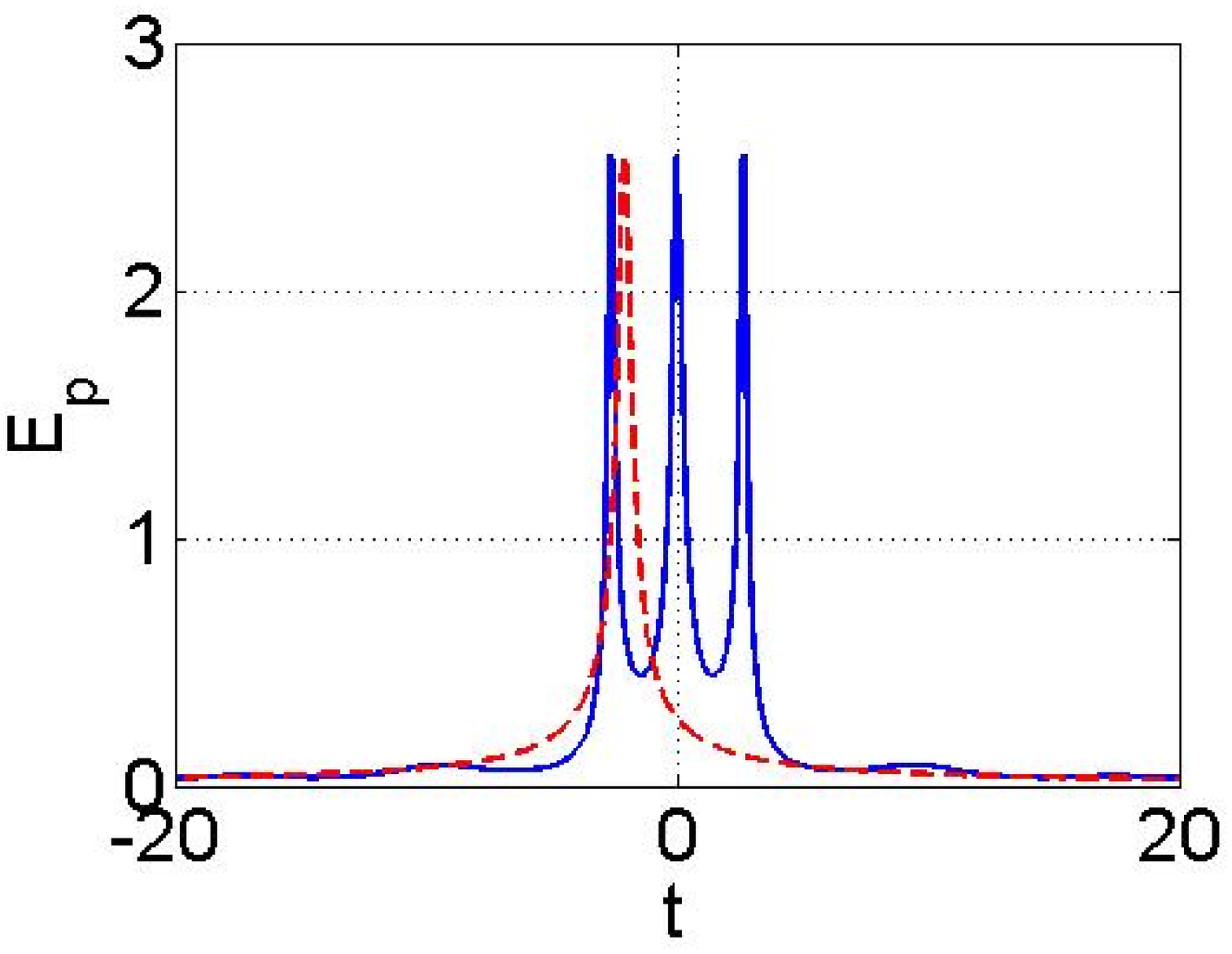}}
\begin{center}
\hskip 1cm $(\rm{a})$ \hskip 8cm $(\rm{b})$
\end{center}
\caption{(a) Shape of the difference between light intensities and CW background of the solution (\ref{s5})
at $x=2.11$ with $d=0$ (dashed line) and $d=3\pi/4$ (solid line);
(b) Shape of the energy pulse given by (\ref{EP}) with $d=0$ (dashed line) and $d=3\pi/4$ (solid line).
The other parameters are the same as depicted in Fig. \ref{fig:2}. }
\label{fig:11}
\end{figure}

At last, let us simply discuss some wave characteristics of the interesting dark rogue wave
with multiple compression points. The difference between light intensity and CW background
of the nonautonomous dark rogue wave solution (\ref{s5}) is defined as
\begin{equation}
\Delta I_{c}(x,t)=|u_{1}[1]|^2-|u_{10}[1]|^2,
\end{equation}
where $u_{10}[1]=\displaystyle\lim_{x\rightarrow\infty}u_{1}[1]$. Hereby, we can check
\begin{equation}
\int_{-\infty}^{+\infty} \Delta I_{c}(x,t){\rm d}t=0,
\end{equation}
which implies that for the dark rogue wave, the energy of the pump is preserved in the fiber with periodic modulation characteristics.
However, it is clearly seen that in Fig. \ref{fig:11}(a)
the light intensity of the nonautonomous dark rogue wave solution is always lower than its CW background,
which is different from the corresponding situation of bright case reported in \cite{35}.

Next, we can obtain the energy pulse of the nonautonomous dark rogue wave defined by
\begin{equation}\label{EP}
E_{p}(t)=\int_{-\infty}^{\infty}|u_{1}[1]-u_{10}[1]|^2{\rm d}t.
\end{equation}
We show that the relative energy pulse is coincide well with the compression points, and
with the amplitude of the periodic modulation increasing, the number of maximum
points of the energy pulse also increases, see Fig. \ref{fig:11}(b).

\section{Conclusion }

In summary, we  presented a hierarchy of rogue wave solutions under two different relative frequencies
based on MI and gDT for the VCCH equations.  An exact calculation formula depending on the amplitude of the periodic
modulation for the number of the compression points of the rogue waves are given.
Thus, the dark-bright and composite rogue waves with multiple compression points,
and even the corresponding comb-like structures are generated by
choosing sufficiently large periodic modulation amplitudes in coefficients of the VCCH equations.
Moreover, for higher-order case, the dark-bright three sisters involving one rogue wave with
triple compression points, the rogue wave quartets involving two rogue waves with triple compression points,
and the rogue wave cluster involving six rogue waves with triple compression points are visually shown by selecting the
adequate modulation coefficients. Further,
some wave wave characteristics of the dark rogue wave solution are discussed in the last of the paper.
Our work can been seen as the vector generalization with higher-order effects
of Tiofack'  very recent results, and it is expected that these findings may provide some
theoretical assistance to the experimental control and manipulation of vector rogue wave dynamics in  inhomogeneous BEC, etc.

\section*{Acknowledgment}
The project is supported by
NNSFC (No. 11275072 and 11435005),
the Global Change Research Program of China (No.2015CB953904),
Doctoral Program of Higher Education of China (No. 20120076110024),
The Network Information Physics Calculation of basic research innovation research group of China (No. 61321064),
and Shanghai Collaborative Innovation Center of Trustworthy Software for Internet of Things (No. ZF1213).

\section*{Appendix:Polynomials}
\begin{align}
&G_{1}=[12 x^2+(4 \sqrt{3} z_{1}+4 z_{1}) x+(756 \nu^2-72 \nu z_{1}+24 \sqrt{3} \nu z_{1}+48 \sqrt{3}) A(t)^2
+((8 \sqrt{3} z_{1}\nonumber \\&~~~~-144 \nu +24 z_{1}) x+360 \sqrt{3} \nu-24 \sqrt{3} \nu z_{1}-24 \nu z_{1}
+40 \sqrt{3} z_{1}+120 z_{1}+24) A(t)+297\nonumber \\&~~~~+\sqrt{3}+\sqrt{3} z_{1}+3 z_{1}],\nonumber\\
&H_{1}=[(6+2 \sqrt{3} z_{1}) x  +(8 \sqrt{3} z_{1}-36 \nu z_{1}-12 \sqrt{3}+18 \nu+12 z_{1}-30 \sqrt{3} \nu z_{1}) A(t)
\nonumber\\&~~~~+(33 \sqrt{3}-29 z_{1}-3-19 \sqrt{3} z_{1})],\nonumber\\
&G_{2}=[12 x^2+(4 \sqrt{3} z_{1}+4 z_{1}) x+(756 \nu^2-72 \nu z_{1}+24 \sqrt{3} \nu z_{1}+48 \sqrt{3}) A(t)^2+((8 \sqrt{3} z_{1}
\nonumber\\&~~~~+24 z_{1}-144 \nu) x
+360 \nu \sqrt{3}-24 \sqrt{3} \nu z_{1}-24 \nu z_{1}+40 \sqrt{3} z_{1}+120 z_{1}+24) A(t)+297\nonumber\\&~~~~+\sqrt{3}-\sqrt{3} z_{1}-3 z_{1} ], \nonumber\\
&H_{2}=[(-6+2 \sqrt{3} z_{1}) x-(30 \sqrt{3} \nu z_{1}+36 \nu z_{1}+12 \sqrt{3}
+8 \sqrt{3} z_{1}+12 z_{1}+18 \nu) A(t)\nonumber\\&~~~~-(21 \sqrt{3} z_{1}+31 z_{1}+27 \sqrt{3}+3)], \nonumber\\
&F=2[3 x^2+(z_{1}+\sqrt{3} z_{1}) x+(189 \nu^2-18 \nu z_{1}+12 \sqrt{3}+6 \sqrt{3} \nu z_{1}) A(t)^2
+((6 z_{1}-36 \nu+2 \sqrt{3} z_{1}) x\nonumber\\&~~~~+10 \sqrt{3} z_{1}+30 z_{1}+90 \sqrt{3} \nu-6 \nu z_{1}
+6-6 \sqrt{3} \nu z_{1}) A(t)+75+\sqrt{3}],  \nonumber\\
&\rho_{1}=\frac{(3+\sqrt{3}z_{1}+2z_{1})}{(6+6{\rm i}+\sqrt{3} z_{1}+z_{1}+3{\rm i} z_{1}+\sqrt{3}{\rm i} z_{1})},\
\rho_{2}=\frac{(-3+\sqrt{3} z_{1}+2 z_{1})}{(-6-6{\rm i}+\sqrt{3} z_{1}+z_{1}+3{\rm i} z_{1}+\sqrt{3}{\rm i} z_{1})},\nonumber\\
&\widetilde{G}_{1}=2304 x^4+4608 \sqrt{3} x^3-932352 x^2-920576 \sqrt{3} x+81 (363 \nu^2+16)^2 A(t)^4
+(-14256 \nu (363 \nu^2\nonumber\\&~~~~+16) x-216 \sqrt{3} (192+16335 \nu^3+4356 \nu^2+1040 \nu)) A(t)^3
+((940896 \nu^2+13824) x^2\nonumber\\&~~~~+288 \sqrt{3} (4059 \nu^2+112+792 \nu) x-124416 \nu-63053856 \nu^2+2797056) A(t)^2
+(-76032 \nu x^3\nonumber\\&~~~~-3456 \sqrt{3} (37 \nu+4) x^2+15342336 \nu x+768 \sqrt{3} (-3576+13507 \nu)) A(t)+93077504,\nonumber\\
&\widetilde{H}_{1}=-2304 \sqrt{3} x^4+4608 x^3+923136 \sqrt{3} x^2-930816 x-81 \sqrt{3} (363 \nu^2+16)^2 A(t)^4
\nonumber\\&~~~~+(14256 \sqrt{3} \nu (363 \nu^2+16) x+235224 \nu^3-41472-940896 \nu^2+217728 \nu) A(t)^3
\nonumber\\&~~~~+(-864 \sqrt{3} (1089 \nu^2+16) x^2+(228096 \nu+256608 \nu^2-41472) x+288 \sqrt{3} (-9488+384 \nu
\nonumber\\&~~~~+218343 \nu^2)) A(t)^2
+(76032 \sqrt{3} \nu x^3-(72576 \nu+13824) x^2-2304 \sqrt{3} (8+6617 \nu) x
\nonumber\\&~~~~-2783232-573696 \nu) A(t)-91238400 \sqrt{3}, \nonumber\\
&\widetilde{G}_{2}=2304 x^4+4608 \sqrt{3} x^3-932352 x^2-920576 \sqrt{3} x+81 (363 \nu^2+16)^2 A(t)^4
+(-14256 \nu (363 \nu^2\nonumber\\&~~~~+16) x-216 \sqrt{3} (-192-4356 \nu^2+1040 \nu+16335 \nu^3)) A(t)^3
+((940896 \nu^2+13824) x^2\nonumber\\&~~~~+288 \sqrt{3} (112+4059 \nu^2-792 \nu) x+124416 \nu-63053856 \nu^2+2797056) A(t)^2
+(-76032 \nu x^3\nonumber\\&~~~~-3456 \sqrt{3} (37 \nu-4) x^2+15342336 \nu x+768 \sqrt{3} (13507 \nu+3576)) A(t)+93077504,\nonumber\\
&\widetilde{H}_{2}=2304 \sqrt{3} x^4-4608 x^3-923136 \sqrt{3} x^2+930816 x+81 \sqrt{3} (363 \nu^2+16)^2 A(t)^4
\nonumber\\&~~~~+(-14256 \sqrt{3} \nu (363 \nu^2+16) x-235224 \nu^3-41472-940896 \nu^2-217728 \nu) A(t)^3\nonumber
\end{align}

\begin{align}
&~~~~+(864 \sqrt{3} (1089 \nu^2+16) x^2
+(228096 \nu-256608 \nu^2+41472) x-288 \sqrt{3} (-9488-384 \nu\nonumber\\&~~~~+218343 \nu^2)) A(t)^2
+(-76032 \sqrt{3} \nu x^3+(72576 \nu-13824) x^2+2304 \sqrt{3} (-8+6617 \nu) x
\nonumber\\&~~~~-2783232+573696 \nu) A(t)+91238400 \sqrt{3},\nonumber\\
&\widetilde{F}=-4608 x^4+1837056 x^2+4096 \sqrt{3} x
-162 (363 \nu^2+16)^2 A(t)^4
+(28512 \nu (363 \nu^2+16) x\nonumber\\&~~~~+1728 \sqrt{3} \nu (128+1089 \nu^2)) A(t)^3
+((-1881792 \nu^2-27648) x^2-4608 \sqrt{3} (99 \nu^2+8) x\nonumber\\&~~~~+124910208 \nu^2-5566464) A(t)^2+
(152064 \nu x^3+27648 \sqrt{3} \nu x^2-30311424 \nu x\nonumber\\&~~~~-5563392 \sqrt{3} \nu) A(t)
-184324096.\nonumber
\end{align}

\end{document}